\documentclass[reprint,aps,prd,amsmath,amssymb,nofootinbib,superscriptaddress,floatfix]{revtex4-2}
\usepackage{graphicx}
\usepackage{dcolumn}
\usepackage[utf8]{inputenc}
\usepackage{xcolor}
\usepackage[unicode=true,colorlinks,linkcolor=blue,anchorcolor=blue,urlcolor=blue,citecolor=blue,breaklinks=true]{hyperref}
\usepackage{epstopdf}
\usepackage{orcidlink}

\newcommand{\pararrow}{\mathord{\buildrel{\lower3pt\hbox{$\scriptscriptstyle\leftrightarrow$}}\over {\partial}}} % partial leftrightarrow operation
\newcommand{\pararrowk}[1]{\mathord{\buildrel{\lower3pt\hbox{$\scriptscriptstyle\leftrightarrow$}}\over {\partial}\hspace*{-0.18em}{}^#1}\hspace*{-0.18em} \,} % partial
\usepackage{bm}
\usepackage{slashed}
\allowdisplaybreaks
\usepackage{multirow}

\newcommand{\tju}{\affiliation{Center for Joint Quantum Studies and Department of Physics, School of Science, Tianjin University, Tianjin 300350, China}}
\newcommand{\qfnu}{\affiliation{College of Physics and Engineering, Qufu Normal University, Qufu 273165, China}}

\begin{document}

\title{Production of hidden-charm molecular candidates in $\psi(4660)$ decays}

\author{Xiao-Yun Wang\,\orcidlink{0009-0006-6091-6944}} \tju
\author{Ya-Rong Wang\,\orcidlink{0000-0003-3158-5577}} \tju
\author{Shi-Dong Liu\,\orcidlink{0000-0001-9404-5418}}~\email{liusd@qfnu.edu.cn} \qfnu
\author{Xiao-Hai Liu\,\orcidlink{0000-0002-6159-3140}}~\email{xiaohai.liu@tju.edu.cn} \tju
\author{Gang Li\,\orcidlink{0000-0002-5227-8296}}~\email{gli@qfnu.edu.cn} \qfnu

\begin{abstract}
We investigate the production of several hidden-charm exotic candidates, including $Z_c(3900)$, $Z_c(4020)$, $Z_{cs}(3985)$, and $Z_2(4250)$, in $\psi(4660)$ decays under the assumption that these states are predominantly hadronic molecules. Treating $\psi(4660)$ as a conventional $\psi(5S)$ charmonium state, the production mechanisms are described through intermediate charmed-meson triangle loops, with its couplings to charmed-meson pairs estimated within the quark model. A systematic analysis of the processes $\psi(4660)\to Z_c(3900)\pi$, $\psi(4660)\to Z_c(4020)\pi$, $\psi(4660)\to Z_{cs}(3985)K$, and $\psi(4660)\to Z_2(4250)\pi$ is performed within a unified framework. The predicted branching fractions are found to be of the order of $10^{-2}$, $10^{-4}$, $10^{-3}$, and $10^{-6}$, respectively, exhibiting only a mild dependence on the cutoff parameter. We further find that the contributions from the $SHH$ intermediate loops dominate over those from the $THH$ and $HHH$ loops in most channels. The sizable production rates obtained in this work indicate that $\psi(4660)$ decays provide a promising platform for probing the molecular nature of charged hidden-charm exotic states and testing their underlying production mechanisms.
\end{abstract}

\maketitle

%%%%%%%%%%%%%%%%%%%%%%%%%%%%%%%%%%
\section{Introduction}
\label{sec:introduction}

Over the past two decades, a large number of exotic hadron states have been observed in the heavy-quark sector. Most of these states cannot be accommodated by the conventional quark model and are collectively referred to as $XYZ$ states. 
Their discoveries have opened a new chapter in hadron spectroscopy and also drawn sustained interest from both the experimental and theoretical communities. 
Understanding their internal structures is essential for revealing the nonperturbative dynamics of quantum chromodynamics (QCD). 
Various interpretations have been proposed, including compact tetraquarks, hadrocharmonium, hybrids, hadronic molecules, and threshold effects. For recent reviews, we refer the reader to Refs.~\cite{Brambilla:2019esw,Chen:2016qju,Guo:2017jvc,Liu:2019zoy,Meng:2022ozq,Bai:2026atm,Wang:2025dur}.

Among the charged charmonium-like states observed so far, $Z_c(3900)$ and $Z_c(4020)$ have attracted particular attention. 
The $Z_c(3900)$ was first observed in 2013 by the BESIII Collaboration in the $J/\psi \pi^{\pm}$ invariant mass spectrum of the process $e^+e^-\to J/\psi \pi^+\pi^-$ at $\sqrt{s}= 4.26$ GeV~\cite{BESIII:2013ris}, and was subsequently confirmed by the Belle Collaboration in the same process~\cite{Belle:2013yex}. 
The charged $Z_c(3900)$ was also observed in the $D\bar D^*$ invariant mass spectrum in the open charm process $e^+e^-\to\pi^\pm(D\bar D^*)^\mp$~\cite{BESIII:2013qmu}. 
Two years later, the BESIII Collaboration observed the neutral $Z_c(3900)$ in the $e^+e^-\to J/\psi \pi^0\pi^0$~\cite{BESIII:2015cld} and $e^+e^- \to \pi^0(D{\bar D}^*)^0$~\cite{BESIII:2015ntl}.

The charged $Z_c(4020)$ state, as the heavy quark spin symmetry partner of $Z_c(3900)$, was found by the BESIII Collaboration in the $h_c \pi^\pm$ invariant mass spectrum of the process $e^+e^-\to h_c\pi^+\pi^-$~\cite{BESIII:2013ouc} and also in the $D^*\bar D^*$ invariant mass spectrum of $e^+e^-\to\pi^\pm(D^*\bar D^*)^\mp$~\cite{BESIII:2013mhi}. 
Its neutral partner was discovered in the $h_c\pi^0$ and $(D^{*}\bar D^{*})^{0}$ invariant mass spectra of the processes $e^+e^-\to h_c \pi^0\pi^0$~\cite{BESIII:2014gnk} and $e^{+}e^{-}\to(D^{*}\bar D^{*})^{0}\pi^0$~\cite{BESIII:2015tix}, respectively. 

Since the measured masses of $Z_c(3900)$ and $Z_c(4020)$ lie slightly above the $D\bar D^*$ and $D^*\bar D^*$ thresholds, respectively, it is natural to regard them as hadronic molecule candidates~\cite{Aceti:2014uea,Guo:2013sya,Cui:2013yva,Zhang:2013aoa,Chen:2013omd}. 
For instance, in Refs.~\cite{Li:2013xia,Li:2014pfa,Chen:2015igx,Chen:2016byt,Xiao:2018kfx}, the production and decay properties of the $Z_c(3900)$ and $Z_c(4020)$ were investigated within the molecular framework.
Besides the molecular interpretation, other theoretical interpretations of the $Z_c(3900)$ and $Z_c(4020)$ are also possible, such as tetraquark states~\cite{Braaten:2013boa,Faccini:2013lda,Wang:2013llv,Qiao:2013dda}, a virtual state~\cite{He:2017lhy}, and kinematic effects~\cite{Wang:2013cya,Liu:2013vfa,Swanson:2014tra,Szczepaniak:2015eza,Yu:2024sqv}. 

According to SU(3) flavor symmetry, a strange partner of the $Z_c$ state with the quark content $c \bar{c} s\bar{q}$ is expected to exist. 
The $Z_{cs}(3985)$, observed in 2020 by the BESIII Collaboration in the recoil-mass spectrum of $K^+$ from the process $e^+e^-\to K^+(D_s^- D^{*0}+D_s^{*-}D^0)$~\cite{Ablikim:2020hsk}, can be regarded as the strange partner of the $Z_{c}(3900)$, thereby further enriching the spectroscopy of charged hidden charm exotic states. 
Subsequently, the LHCb Collaboration observed a structure $Z_{cs}(4000)$ in the decay $B^+ \to J/\psi \phi K^+$~\cite{Aaij:2021ivw}. 
The mass and width of $Z_{cs}(3985)$ are $(3982.5^{+1.8}_{-2.6}\pm 2.1)$ MeV and $(12.8^{+5.3}_{-4.4}\pm 3.0)$ MeV, respectively, while those of $Z_{cs}(4000)$ are $(4003\pm 6^{+4}_{-14})$ MeV and $(131\pm 15\pm 26)$ MeV. 
Whether the $Z_{cs}(3985)$ and $Z_{cs}(4000)$ are the same state remains an open question~\cite{Chen:2022asf,Du:2022jjv,Yang:2020nrt,Ortega:2021enc,Meng:2021rdg,Ikeno:2021mcb}. 
In this work, following Ref.~\cite{Ortega:2021enc}, we assume they are the same state and refer to it as $Z_{cs}$.
Owing to the proximity of the $Z_{cs}$ to the $D_s\bar D^*/D_s^*\bar D$ thresholds, it can be considered as a hadronic molecule candidate~\cite{Yang:2020nrt,Meng:2020ihj,Du:2020vwb,Yan:2021tcp}. 
The $Z_{cs}$ can also be interpreted either as a tetraquark state or as a kinematic effect, as discussed in Refs.~\cite{Wang:2020iqt,Wan:2020oxt,Wang:2020kej,Yang:2020nrt,2103.05282}.

Beyond the near-threshold candidates composed of ground-state mesons, it is also important to explore possible molecular configurations with higher masses and unusual quantum numbers. The Belle Collaboration observed two resonancelike structures $Z_1^+(4051)$ and $Z_2^+(4250)$ in the $\pi^+ \chi_{c1}$ invariant mass spectrum of the process $\bar{B}^0 \to K^- \pi^+ \chi_{c1}$~\cite{Belle:2008qeq}. Their masses and widths are determined to be $M_1 = (4051 \pm 14^{+20}_{-41})$ MeV, $\Gamma_1 = (82^{+21+47}_{-17-22})$ MeV and $M_2= (4248^{+44+180}_{-29-35})$ MeV, $\Gamma_2=(177^{+54+316}_{-39-61})$ MeV, respectively. The Particle Data Group (PDG) now labels $Z_2^+(4250)$ as $T_{c\bar c}^+(4250)$~\cite{ParticleDataGroup:2024cfk}. Some studies suggest that the $Z_2(4250)$ could be interpreted as the $D_1\bar D$ molecule with $J^P=1^-$~\cite{Ding:2008gr,Lee:2008tz}. Compared to the lower $Z_c$ states, the properties of such a $1^-$ molecular state would be more sensitive to the dynamics of intermediate excited mesons.

To better understand the nature of exotic states, it is instructive to explore their production mechanisms. The $Y$ states, such as the vector charmonium-like state $\psi(4660)$, serve as an effective source for producing hidden-charm molecular states in association with a pion or a kaon.
The $\psi(4660)$ was first observed by the Belle Collaboration in the $\psi(2S)\pi^+\pi^-$ invariant mass spectrum via initial state radiation processes~\cite{Wang:2007ea}. 
Its mass and width are approximately $(4664 \pm 11\pm 5)$ MeV and $(48\pm 15\pm 3)$ MeV. 
In Ref.~\cite{Ding:2007rg}, the $\psi(4660)$ was suggested to be identified as a $5^3S_1$ $c\bar{c}$ candidate, and its $e^+e^-$ leptonic widths, $E1$ and $M1$ transitions, as well as open-flavor strong decays were evaluated accordingly.
However, within the screened potential model~\cite{Li:2009zu}, the $Y(4660)$ was assigned to be the $\psi(6S)$ charmonium state.
In addition to these charmonium interpretations, several alternative pictures have been put forward, including the $f_0(980)\psi^\prime$ molecule~\cite{Guo:2008zg,Guo:2009id,Albuquerque:2011ix}, a hadronic state linked to $\Lambda_c\bar{\Lambda}_c$ dynamics~\cite{Ding:2007rg,Badalian:2008dv,Simonov:2008cr,Dai:2017fwx,Ahmad:2025mue}, and a tetraquark configuration~\cite{Maiani:2014aja,Ebert:2008kb,Chen:2010ze}.
Analyses based on QCD sum rules point to tetraquark configurations such as $qc\bar{q}\bar{c}$ or $sc\bar{s}\bar{c}$~\cite{Chen:2010ze}.
Notably, the $\psi(4660)$ mass lies above several open-charm thresholds, allowing it to couple to a variety of charmed meson pairs and providing a favorable environment for investigating rescattering effects induced by intermediate charmed meson loops. 
In this work, we treat the $\psi(4660)$ as a pure $\psi(5S)$ state.

This paper is organized as follows. In Sec.~\ref{sec:framework}, we present the effective Lagrangians, vertex functions, and rescattering amplitudes used in our calculation. In Sec.~\ref{sec:numerical}, we show the numerical results and discuss the dependence of the branching ratios on the cutoff parameter, as well as the relative importance of different intermediate channels. Finally, a summary is given in Sec.~\ref{sec:summary}.

\section{Theoretical Framework}
\label{sec:framework}

\subsection{Vertex functions}\label{Lagrangian}

We describe the interactions of heavy mesons containing a single heavy quark within the framework of heavy quark effective theory (HQET), which exploits the heavy quark spin and flavor symmetries in the limit $m_Q \to \infty$. In this limit, the heavy quark four-velocity $v$ coincides with that of the hadron and is conserved in strong interactions. Owing to heavy quark spin symmetry, hadrons that differ only in the orientation of the heavy quark spin are degenerate in mass and form spin doublets. 

For the ground-state charmed mesons with orbital angular momentum $l=0$ for the light degrees of freedom, the heavy quark spin doublet has negative parity spin quantum numbers $J^P = (0^-, 1^-)$ and corresponds to $(D_{(s)}, D_{(s)}^*)$. The doublet can be represented by the $4\times 4$ superfield 
\begin{equation}
H_{a} = \frac{1 + \slashed{v}}{2} \left[ \mathcal{D}_{a\mu}^* \gamma^\mu - \mathcal{D}_a \gamma_5 \right],
\end{equation}
where $a$ is the light-flavor index. The symbols $\mathcal{D}_a$ and $\mathcal{D}_{a\mu}^*$ denote the pseudoscalar and vector charmed mesons, respectively, i.e., $\mathcal{D} = (D^0, D^+, D_s^+)$ and $\mathcal{D}_{\mu}^* = (D^{*0}, D^{*+}, D_s^{*+})$.

For the $P$-wave ($l=1$) heavy-light mesons, HQET predicts two distinct doublets. One contains the states with $J^P=0^+$ and $J^P=1^+$ whose corresponding superfield is
\begin{align}
S_{a} &= \frac{1+\slashed{v}}{2} [{\cal D}_{1a}^{\prime \mu} \gamma_{\mu} {\gamma}_5 - {\cal D}_{0a}].
\end{align}
The other consists of the states with $J^P=1^+$ and $J^P=2^+$, combined into the superfield 
\begin{align}
T_{a}^{\mu} &=  \frac{1+\slashed{v}}{2} \big[{\mathcal D}_{2a}^{\mu \nu} {\gamma}_{\nu} - \sqrt{\frac{3}{2}} {\mathcal D}_{1a\nu} {\gamma}_5 [g^{\mu \nu} - \frac{1}{3} {\gamma}^{\nu} ({\gamma}^{\mu} - v^{\mu})] \big].
\end{align}

\begin{figure}
\centering
\includegraphics[width=1.0\hsize]{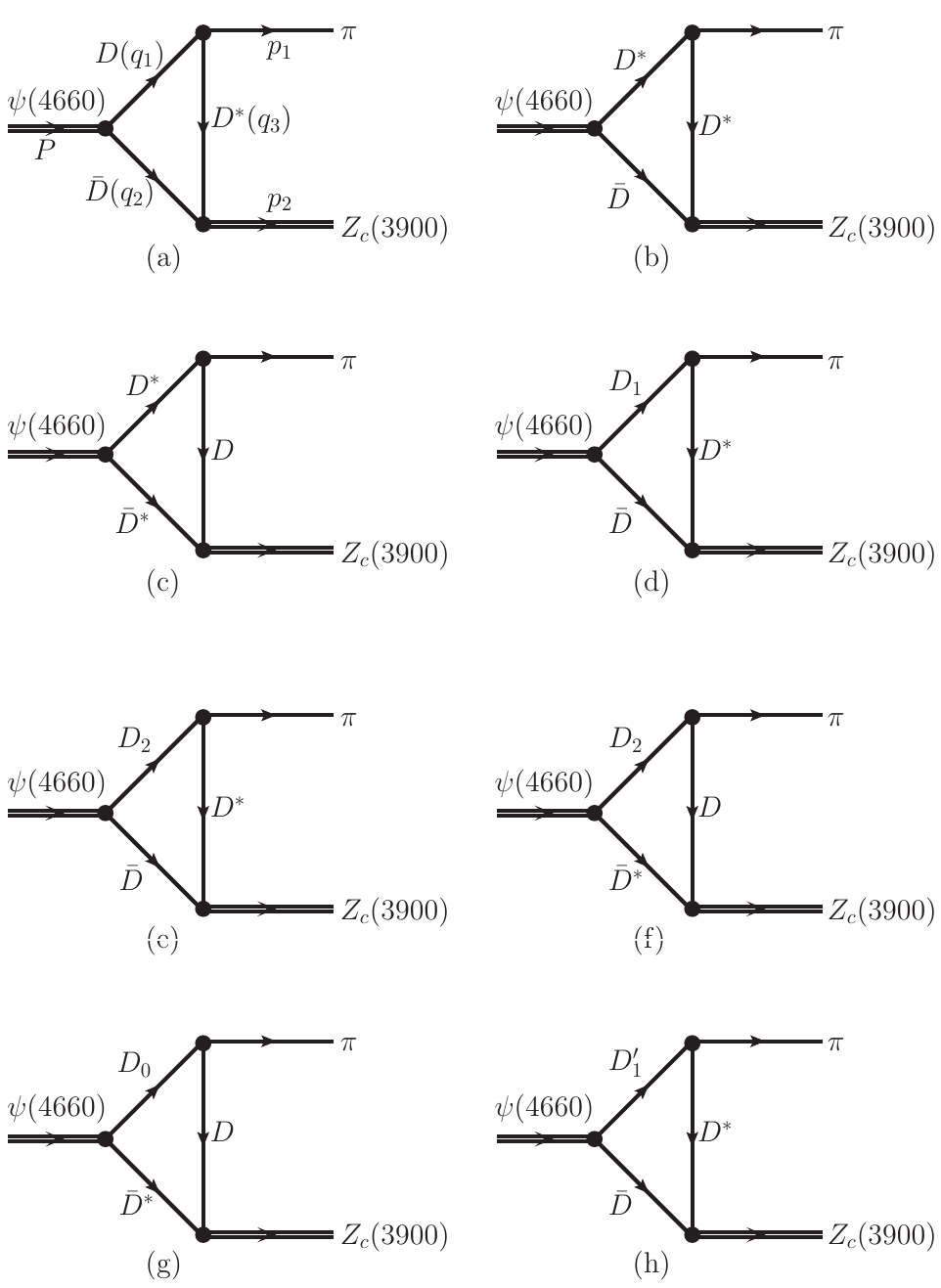}
\caption{Triangle diagrams for $\psi(4660) \to Z_c(3900) \pi$: 
(a)--(c) $HHH$ loops; 
(d)--(f) $THH$ loops; 
(g) and (h) $SHH$ loops. The relevant kinematics [$P, p_{1(2)}, q_{1(2,3)}$] are explicitly indicated in the graph (a).}
\label{fig:triangle Zc}
\end{figure}

\begin{figure}[htbp]
\centering
\includegraphics[width=1.0\hsize]{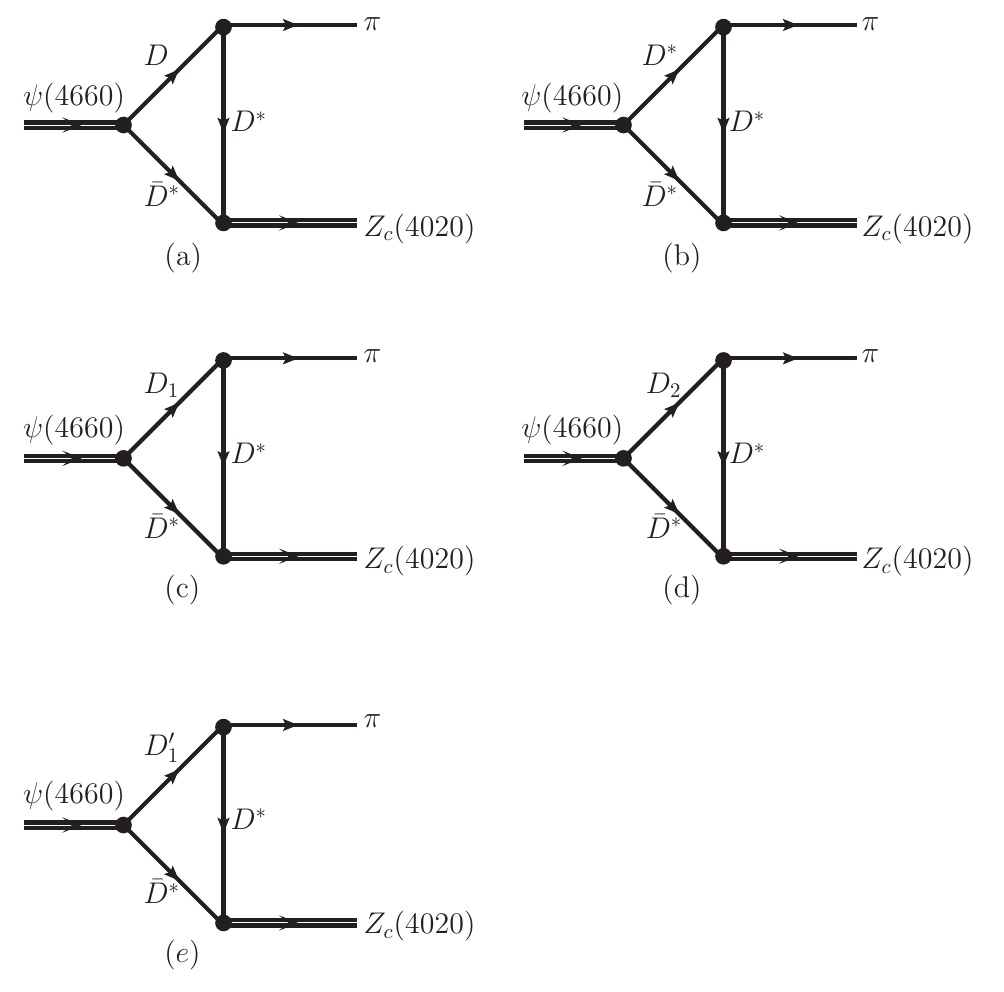}
\caption{Triangle diagrams for $\psi(4660) \to Z_c(4020) \pi$: 
(a) and (b) $HHH$ loops; 
(c) and (d) $THH$ loops; 
(e) $SHH$ loop.}
\label{fig:triangle Zcprime}
\end{figure}

\begin{figure}
\centering
\includegraphics[width=1.0\hsize]{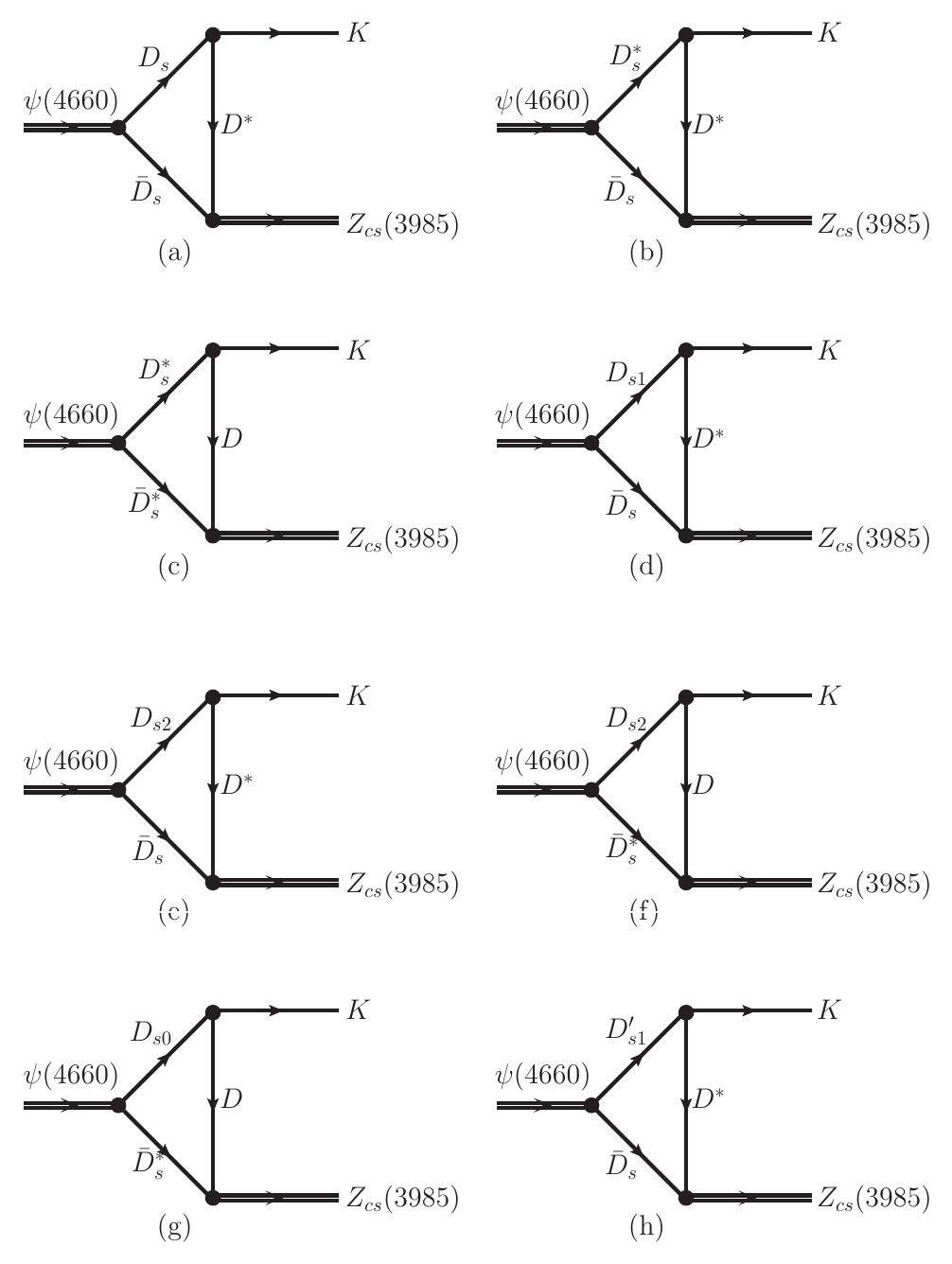}
\caption{Triangle diagrams for $\psi(4660) \to Z_{cs}(3985) K$: 
(a)(b)(c) $HHH$ loop; 
(d)(e)(f) $THH$ loop; 
(g)(h) $SHH$ loop.}
\label{fig:triangle Zcs}
\end{figure}

\begin{figure}
\centering
\includegraphics[width=1.0\hsize]{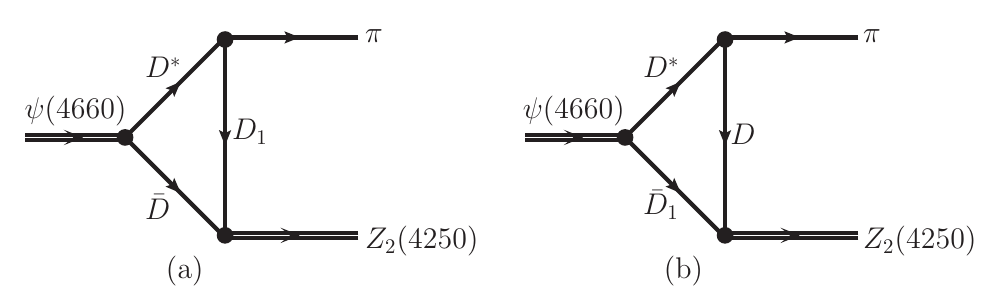}
\caption{Triangle diagrams for $\psi(4660) \to Z_2(4250) \pi$: (a) $D^*\bar{D}D_1$ loop; (b) $D^*\bar{D}_1 D$ loop.}
\label{fig:triangle Z2}
\end{figure}

The triangle diagrams considered in this work for $\psi(4660) \to Z_c(3900) \pi$ and $\psi(4660) \to Z_c(4020) \pi$ are shown in Figs.~\ref{fig:triangle Zc} and \ref{fig:triangle Zcprime}, respectively. In addition, the triangle diagrams for $\psi(4660) \to Z_{cs}(3985) K$ and $\psi(4660) \to Z_2(4250) \pi$ are presented in Figs.~\ref{fig:triangle Zcs} and \ref{fig:triangle Z2}, respectively.
In the following, we 
consider the $Z_c(3900)$ and $Z_c(4020)$ as the molecular state made of $D^* \bar D(D \bar D^*)$ and $D^* \bar D^*$, respectively, while the $Z_{cs}(3985)$ and $Z_2(4250)$ are assumed to be 
$D^* D_s/ D^*_s D$ and $D_1 \bar D$ molecular states.

To evaluate the leading contributions from the charmed meson loops, we employ the leading order effective Lagrangians constrained by the heavy quark symmetry and chiral symmetry~\cite{Casalbuoni:1996pg}. The relevant effective Lagrangian reads
\begin{eqnarray}
&& {\mathcal L}_{P} = i g \langle H_{b} \gamma_\mu \gamma_5 {\cal A}^\mu_{ba} \bar{H}_{a} \rangle  + i h \langle S_{b} \gamma_\mu \gamma_5 {\cal A}_{ba}^\mu {\bar H}_{a} \rangle,\nonumber \\
&& + i\frac{h^\prime}{\Lambda_\chi} \langle \bar{H}_{a} T_{b}^\mu (i \mathcal{D}_\mu \slashed{\cal A} + i \slashed{\cal D} {\cal A}_\mu)_{ba} \gamma_5 \rangle + \mathrm{H.c.},\label{eq:Lag}
\end{eqnarray}
with
\begin{eqnarray}
&&{\bar H}_{a}= {\gamma}^0 H_{a}^{\dagger} {\gamma}^0, 
\end{eqnarray}
where $\langle \cdots \rangle$ denotes the trace over Dirac matrices. ${\mathcal A}^\mu$ is the chiral axial vector containing the Goldstone bosons, defined as
\begin{eqnarray}
{\mathcal A}_\mu = \frac{1}{2}(\xi^\dagger \partial_\mu \xi - \xi \partial_\mu \xi^\dagger),
\end{eqnarray}
with $\xi = e^{i{\mathcal M}/f_{\pi}}$ and
\begin{eqnarray}
&&{\mathcal M} = \left( 
\begin{array}{ccc} 
\frac{1}{\sqrt{2}}\pi^0 + \frac{1}{\sqrt{6}}\eta & \pi^+ &K^+ \\
\pi^- & -\frac{1}{\sqrt{2}}\pi^0 + \frac{1}{\sqrt{6}}\eta & K^0 \\
K^- & {\bar K}^0 & -\sqrt{\frac{2}{3}}\eta 
\end{array} 
\right).
\end{eqnarray}

According to Eq.~(\ref{eq:Lag}), the vertex functions $\mathcal{A}_{H/S/T(q_1) \to H(q_3) \pi(p_1)}$ in Fig.~\ref{fig:triangle Zc} are
\begin{align}
&\mathcal{A}_{D\to D^*\pi} = -g_{D^*D\mathcal{M}} p_1^{\mu} \varepsilon_\mu(D^*), \\
&\mathcal{A}_{D^*\to D^*\pi} = ig_{D^*D^* \mathcal{M}}\epsilon^{\mu \nu \alpha \beta} v_{\nu} p_{1\mu} \varepsilon_\alpha(D^*) \varepsilon_{\beta}(D^*),\\
&\mathcal{A}_{D_0\to D\pi} = -g_{D D_0 \mathcal{M}}p_{1\mu} v^{\mu},\\
&\mathcal{A}_{D_1^{\prime}\to D^*\pi} = -g_{D^*D_1^{\prime}\mathcal{M}}p_{1\mu} v^{\mu} \varepsilon_{\nu}(D_1^{\prime}) \varepsilon^{\nu}(D^*),\\
&\mathcal{A}_{D_1\to D^*\pi} = -ig_{D^*D_1\mathcal{M}} (3 p_1^{\mu} p_1^{\nu}-p_1^2 g^{\mu \nu}) \varepsilon_\mu(D_1) \varepsilon_\nu(D^*),\\
&\mathcal{A}_{D_2\to D^*\pi} =  g_{D^*D_2\mathcal{M}}\epsilon^{\mu \nu \alpha \beta} v_\alpha p_{1 \mu} p_1^{\eta} \varepsilon_{\eta \nu}(D_2) \varepsilon_\beta(D^*),\\
&\mathcal{A}_{D_2\to D\pi} = -ig_{DD_2\mathcal{M}} p_1^{\mu} p_1^{\nu} \varepsilon_{\mu \nu}(D_2) .
\end{align}

The interactions between the $Z_c(3900)$/$Z_c(4020)$ and their constituents, $D^* \bar D(D \bar D^*)$/$D^* \bar D^*$ can be described by the effective Lagrangians
\begin{eqnarray}
{\mathcal{L}}_{Z_c D \bar D^*} &=& g_{Z_c} Z_c^{\mu} (D \bar D^*_{\mu} + D^*_{\mu} \bar D),\label{eq:Zc}\\
{\mathcal{L}}_{Z^{\prime}_c D^* \bar D^*} &=& i g_{Z^{\prime}_c} \epsilon_{\mu \nu \alpha \beta} \partial^{\mu}Z_c^{\prime \nu} D^{* \alpha} \bar D^{* \beta},\label{eq:Zcp}
\end{eqnarray}
where we use $Z_c$ and $Z_c^\prime$ to indicate $Z_c(3900)$ and $Z_c(4020)$, respectively.

Assuming the $Z_{cs}(3985)$ to be a $D^* D_s/ D^*_s D$ molecular state, the corresponding effective Lagrangian can be written as 
\begin{eqnarray}
    \mathcal{L}_{Z_{cs} D_s D^*} &=& g_{Z_{cs}} Z_{cs}^{\mu} \left( D^*_{\mu} \bar D_s + D \bar D_{s \mu}^* \right).
    \label{eq:Zcs}
\end{eqnarray}

The $Z_2^+(4250)$ is also a promising molecular candidate with $J^P=1^-$ and is assumed to be dominated by the $D_1 \bar D$ component. Its flavor wave function can be written as
\begin{align}
&&|Z_2^+(4250)\rangle = \frac{1}{\sqrt{2}} \left[ |D_1^+\bar{D}^0\rangle +|D^+ \bar{D}_1^0\rangle \right].
\end{align}
The corresponding effective Lagrangian is 
\begin{align}
&&{\mathcal L}_{Z_2^+(4250) D_1 D}= \frac{g_{Z_2}}{\sqrt{2}} Z_2^{\mu\dagger}(D_{1\mu}^{+} \bar{D}^0 + D^+ \bar{D}_{1\mu}^{0}). \label{eq:Z2}
\end{align}

 The couplings between charmonium and charmed mesons can be constructed within the covariant tensor formalism~\cite{Zou:2002ar,Dulat:2005in,Huang:2026egv}. In this approach, the partial-wave amplitudes are built from the pure orbital-angular-momentum covariant tensors, the metric tensor, the totally antisymmetric Levi-Civita tensor, and the four-momenta of the participating particles. For a process $a\to bc$, the covariant tensor $\tilde{t}^{(l)}_{\mu_1\cdots \mu_l}$ for a final state with pure orbital angular momentum $l$ can be written as
\begin{eqnarray}
    \tilde{t}^{(0)} &=& 1,\\
    \tilde{t}^{(1)}_\mu &=& \tilde{g}_{\mu \nu}(p_a) r^{\nu}\equiv \tilde{r}_\mu,\\
    \tilde{t}^{(2)}_{\mu\nu} &=& \tilde{r}_{\mu} \tilde{r}_{\nu} - \frac{1}{3} (\tilde{r} \cdot \tilde{r}) \tilde{g}_{\mu \nu}(p_a)
\end{eqnarray}
with $\tilde{g}^{\mu\nu}(p_a)=g^{\mu\nu}- p_a^{\mu}p_a^{\nu}/p_a^2$ and $r=p_b-p_c$. For a meson $a$ with spin $S$ and the corresponding spin wave function $\phi^{\mu_1\cdots\mu_S}(p_a,m)$, it is usually necessary to introduce the spin projection operator when constructing the decay amplitudes,
\begin{align}
  &{\mathcal P}^{(1)}_{\mu\mu{^\prime}}(p_a) = \sum_{m} \phi_{\mu}(p_a,m) \phi^*_{\mu^{\prime}}(p_a,m) = -\tilde{g}^{\mu\nu}(p_a),\\
  &{\mathcal P}^{(2)}_{\mu\nu\mu{^\prime}\nu{^\prime}}(p_a) = \frac{1}{2}(\tilde{g}_{\mu\mu^\prime}\tilde{g}_{\nu\nu^\prime} + \tilde{g}_{\mu\nu^\prime}\tilde{g}_{\nu\mu^\prime})-\frac{1}{3}\tilde{g}_{\mu\nu}\tilde{g}_{\mu^\prime \nu^\prime},\\
  &{\mathcal P}^{(3)}_{\mu\nu\lambda\mu{^\prime}\nu{^\prime}\lambda{^\prime}}(p_a) = -\frac{1}{6}(\tilde{g}_{\mu\mu^\prime}\tilde{g}_{\nu\nu^\prime}\tilde{g}_{\lambda\lambda^\prime} + \tilde{g}_{\mu\mu^\prime}\tilde{g}_{\nu\lambda^\prime}\tilde{g}_{\lambda\nu^\prime} \nonumber \\
  &+ \tilde{g}_{\mu\nu^\prime}\tilde{g}_{\nu\mu^\prime}\tilde{g}_{\lambda\lambda^\prime} + \tilde{g}_{\mu\nu^\prime}\tilde{g}_{\nu\lambda^\prime}\tilde{g}_{\lambda\mu^\prime} + \tilde{g}_{\mu\lambda^\prime}\tilde{g}_{\nu\nu^\prime}\tilde{g}_{\lambda\mu^\prime} \nonumber\\
  &+ \tilde{g}_{\mu\lambda^\prime}\tilde{g}_{\nu\mu^\prime}\tilde{g}_{\lambda\nu^\prime}) + \frac{1}{15}(\tilde{g}_{\mu\nu}\tilde{g}_{\mu^\prime\nu^\prime}\tilde{g}_{\lambda\lambda^\prime} +\tilde{g}_{\mu\nu}\tilde{g}_{\nu^\prime\lambda^\prime}\tilde{g}_{\lambda\mu^\prime} \nonumber\\
  &+ \tilde{g}_{\mu\nu}\tilde{g}_{\mu^\prime\lambda^\prime}\tilde{g}_{\lambda\nu^\prime} +\tilde{g}_{\mu\lambda}\tilde{g}_{\mu^\prime\nu^\prime}\tilde{g}_{\nu\lambda^\prime} + \tilde{g}_{\mu\lambda}\tilde{g}_{\nu^\prime\lambda^\prime}\tilde{g}_{\nu\mu^\prime} \nonumber\\
  &+  \tilde{g}_{\nu\lambda}\tilde{g}_{\nu^\prime\lambda^\prime}\tilde{g}_{\mu\mu^\prime} + \tilde{g}_{\nu\lambda}\tilde{g}_{\mu^\prime\nu^\prime}\tilde{g}_{\mu\lambda^\prime} + \tilde{g}_{\nu\lambda^\prime}\tilde{g}_{\mu^\prime\lambda^\prime}\tilde{g}_{\mu\nu^\prime}).
\end{align}

According to conservation of total angular momentum, 
\begin{align}
    J_a=S_{bc}+L_{bc}, \label{eq:angular momentum}\\
    S_{bc}=S_b+S_c.
\end{align}
In addition, parity should be conserved,
\begin{align}
    \eta_a=\eta_b \eta_c (-1)^{L_{bc}},
\end{align}
where $\eta_a$, $\eta_b$, and $\eta_c$ are the intrinsic parities.
This relation determines whether $L_{bc}$ should be even or odd. Then from Eq.~(\ref{eq:angular momentum}), the number of distinct $(L_{bc}, S_{bc})$ combinations can be deduced, which determines the number of independent couplings. It is noted that if $S_{bc} +L_{bc} + J_a$ is odd, the term $i\epsilon_{\mu \nu \alpha \beta} p_a^\beta$ is required; otherwise it is not. In the following amplitudes, different combinations of $(L,S)$ are considered. 

\begin{table*}
\renewcommand\arraystretch{1.5}
\caption{Effective coupling constants for the $\psi SH$, $\psi HH$, and $\psi TH$ vertices, obtained using the MGI model and the QPC model.}
\label{tab:coupling}
\begin{ruledtabular}
\begin{tabular}{cllllllll}
Vertex &Coupling & Value & Coupling & Value & Coupling & Value & Coupling & Value\\
\colrule
\multirow{2}{*}[-0.6ex]{$\psi HH$}
& $g_{\psi D D}$&$0.696$ &$g_{\psi D^* D}$ & $0.178$ GeV$^{-1}$ & $g_{\psi D^* D^*}$&$0.0685$ & $g_{\psi D^* D^*}^{\prime}$ & $0.227$\\
~& $g_{\psi D_s D_s}$&$0.366$ & $g_{\psi D_s^* D_s}$ & $0.0115$ GeV$^{-1}$&$g_{\psi D_s^* D_s^*}$&$0.105$ & $g_{\psi D_s^* D_s^*}^{\prime}$ & $0.334$\\
\colrule
\multirow{4}{*}[-0.6ex]{$\psi TH$}
& $g_{\psi D_1 D}$&$0.483$ GeV$^{-1}$& $g_{\psi D_1 D^*}$ & $0.145$ GeV$^{-2}$& $g_{\psi D_1 D^*}^{\prime}$ & $0.282$ GeV$^{-2}$&$g_{\psi D_2 D}$&$0.00727$ GeV$^{-2}$\\
 & $g_{\psi D_2 D^*}$ & $0.648$ GeV$^{-1}$ & $g_{\psi D_2 D^*}^{\prime}$ & $0.0174$ GeV$^{-3}$ & $g_{\psi D_2 D^*}^{\prime\prime}$ & $2.95$ GeV$^{-1}$\\
 & $g_{\psi D_{s1} D_s}$&$0.645$ GeV$^{-1}$& $g_{\psi D_{s1} D_s^*}$ & $0.589$ GeV$^{-2}$& $g_{\psi D_{s1} D_s^*}^{\prime}$ & $1.18$ GeV$^{-2}$&$g_{\psi D_{s2} D_s}$&$0.139$ GeV$^{-2}$\\
 & $g_{\psi D_{s2} D_s^*}$ & $1.83$ GeV$^{-1}$& $g_{\psi D_{s2} D_s^*}^{\prime}$ & $0.0466$ GeV$^{-3}$& $g_{\psi D_{s2} D_s^*}^{\prime\prime}$ & $8.30$ GeV$^{-1}$\\
\colrule
\multirow{2}{*}[-0.6ex]{$\psi SH$}
 & $g_{\psi D_0 D^*}$ &$1.55$ GeV& $g_{\psi D_1^{\prime} D}$& $0.358$ GeV& $g_{\psi D_1^{\prime} D^*}$&$0.108$\\
 & $g_{\psi D_{s0} D_s^*}$&$1.82$ GeV& $g_{\psi D_{s1}^{\prime} D_s}$& $0.248$ GeV& $$&$$
\end{tabular}
\end{ruledtabular}
\end{table*} 

Regarding the coupling of $\psi(4660)$ to the $HH$ channel, the transition proceeds exclusively through a $P$ wave. Consequently, the corresponding decay amplitudes read
\begin{align}
    \mathcal{A}_{\psi \to DD} &= ig_{\psi  DD} \tilde{t}^{(1)} \cdot \varepsilon(\psi), \label{eq:psiDD}\\
\mathcal{A}_{\psi \to D^*D} &= - g_{\psi D^*D} \epsilon^{\mu \nu \lambda \sigma} \tilde{t}^{(1)}_{\lambda} P_\sigma \varepsilon_{\mu}(\psi) \varepsilon_{\nu}(D^*),\\
\mathcal{A}_{\psi \to D^*D^*} &=i g_{\psi D^* D^*} \tilde{t}^{(1)}_{\mu} \varepsilon^{\mu}(\psi) {\mathcal P}_{\alpha,\beta}^{(1)} \varepsilon^{\beta}(D^*) \varepsilon^{\alpha}(\bar D^*) \nonumber\\
+& ig_{\psi D^* D^*}^{\prime}  \tilde{t}^{(1)\nu} \varepsilon^{\mu}(\psi) {\mathcal P}^{(2)}_{\mu\nu,\alpha\beta}\varepsilon^{\alpha}(D^*) \varepsilon^{\beta}(\bar D^*).
\end{align}
Here $P$ denotes the four-momentum of the $\psi(4660)$. 
In contrast, the couplings of $\psi(4660)$ to the $SH$ channel proceed through an $S$ wave and can be parameterized in a similar way
\begin{align}
\mathcal{A}_{\psi \to D_0 D^*} &= ig_{\psi D_0 D^*} \varepsilon(\psi) \cdot \varepsilon(\bar D^*),\\
\mathcal{A}_{\psi \to D_1^{\prime} D} &= ig_{\psi D_1^{\prime} D} \varepsilon(\psi) \cdot \varepsilon(D_1^{\prime}),\\
\mathcal{A}_{\psi \to D_1^{\prime} D^*} &= - g_{\psi D_1^{\prime} D^*} \epsilon^{\mu \nu \lambda \sigma} P_{\sigma} \varepsilon_\mu(\psi) \varepsilon_\nu(D_1^{\prime}) \varepsilon_\lambda(\bar D^*).
\end{align}

To satisfy the heavy-quark spin symmetry constraints, the couplings of $\psi(4660)$, treated as the conventional $\psi(5S)$ charmonium state, to the $TH$ channels are required to occur through a relative $D$ wave. The corresponding transition amplitudes can be expressed in the covariant tensor formalism as
\begin{align}
  \mathcal{A}_{\psi \to D_1 D} =& ig_{\psi D_1 D} \tilde{t}^{(2)}_{\mu\nu} \varepsilon^{\mu}(\psi) \varepsilon^{\nu}(D_1),\\
\mathcal{A}_{\psi \to D_1 D^*} =& - g_{\psi D_1 D^*} \tilde{t}^{(2)}_{\mu\nu} \epsilon^{\nu \alpha \beta \sigma}  P_{\sigma} \varepsilon^{\mu}(\psi) \varepsilon_{\alpha}(D_1) \varepsilon_{\beta}(\bar D^*) \nonumber \\
-& g_{\psi D_1 D^*}^{\prime} \epsilon^{\mu \alpha \beta \sigma} P_{\sigma} \tilde{t}^{(2)\nu}_{\alpha} \varepsilon_{\mu}(\psi) {\mathcal P}^{(2)}_{\beta\nu,\lambda\rho} \varepsilon^{\lambda}(D_1) \nonumber\\
\times&\varepsilon^{\rho}(\bar D^*), \\
 \mathcal{A}_{\psi \to D_2 D} =& - g_{\psi D_2 D} \epsilon^{\mu \alpha \beta \sigma} \tilde{t}^{(2)}_{\alpha\nu} P_{\sigma} \varepsilon_{\mu}(\psi) \varepsilon_{\beta}^{\nu}(D_2),\\
\mathcal{A}_{\psi \to D_2 D^*} =& ig_{\psi D_2 D^*} \tilde{t}^{(2)\mu\nu} {\mathcal P}^{(1)}_{\nu,\alpha} \varepsilon_{\mu}(\psi) \varepsilon^{\alpha\beta}(D_2) \varepsilon_{\beta}(\bar D^*) \nonumber \\
-& ig_{\psi D_2 D^*}^{\prime}  \epsilon^{\mu \alpha \beta \sigma}P_{\sigma} \varepsilon_{\mu}(\psi) \tilde{t}^{(2)\nu}_{\alpha} {\mathcal P}^{(2)}_{\beta\nu,\lambda\rho}\nonumber\\
\times& \big[ \epsilon^{\lambda \alpha_1 \beta_1 \delta} P_{\delta}  \varepsilon^{\rho}_{ \alpha_1}(D_2) \varepsilon_{\beta_1}(\bar D^*) + \epsilon^{\rho \alpha_2 \beta_2 \delta} P_{\delta} \nonumber\\
\times& \varepsilon^{\lambda}_{ \alpha_2}(D_2) \varepsilon_{\beta_2}(\bar D^*) \big] + i g_{\psi D_2 D^*}^{\prime\prime} \varepsilon^{\mu}(\psi) \tilde{t}^{(2)\nu\lambda} \nonumber\\
\times& \mathcal{P}^{(3)}_{\mu\nu\lambda,\alpha\beta\lambda_1} \varepsilon^{\alpha\beta}(D_2) \varepsilon^{\lambda_1}(\bar D^*).\label{eq:psiD_2D*}
\end{align}

\subsection{Rescattering amplitude}

As an explicit example, the rescattering amplitude corresponding to Fig.~\ref{fig:triangle Zc}(a) for the process $\psi(4660)\to Z_c(3900) \pi$ is given by
\begin{align}\label{eq:ADDD*}
\mathcal{A}_{D D D^*} &= -i\int \frac{d^4 q_3}{(2\pi)^4} \mathcal{A}_{\psi \to D D}  \mathcal{A}_{D \to D^{*} \pi}  \mathcal{A}_{D D^{*} \to Z_c}\nonumber\\
&\times\frac{1}{(q_1^2 - m_{D}^2)(q_2^2 - m_{D}^2)(q_3^2 - m_{D^*}^2)} \mathcal{F}(q_3^2). 
\end{align}
The rescattering amplitudes for the other diagrams in Fig.~\ref{fig:triangle Zc} can be written analogously. The amplitudes for the $\psi(4660)\to Z_c(4020)\pi$, associated with the diagrams in Fig.~\ref{fig:triangle Zcprime}, and those for the $\psi(4660)\to Z_{cs}K$ and $\psi(4660)\to Z_2(4250)\pi$ in Figs.~\ref{fig:triangle Zcs} and \ref{fig:triangle Z2}, are constructed in the same manner. In Eq.~(\ref{eq:ADDD*}), the summation over the polarizations of the intermediate vector and tensor mesons is understood implicitly. For an intermediate spin-1 heavy-meson field, the polarization sum is taken as
\begin{equation}
\sum_{\mathrm{pol}} \epsilon_\mu \epsilon_\nu^* = -g_{\mu\nu}+v_\mu v_\nu.
\end{equation}
Throughout this work, we adopt the nonrelativistic approximation with $v=(1,\boldsymbol{0})$.

To account for the off-shell effects of the exchanged meson and to regularize the ultraviolet behavior of the loop integral, we introduce a monopole form factor~\cite{Gortchakov:1995im,Cheng:2004ru,Wu:2019vbk,Guo:2010ak},
\begin{align}
\mathcal{F}(q_3^2) = \frac{m_3^2 - \Lambda^2}{q_3^2 - \Lambda^2},
\label{eq:F_q}
\end{align}
where $m_3$ is the mass of the exchanged meson and $\Lambda$ denotes the cutoff parameter.

To ensure the self-consistency of the theoretical framework, the definitions and values of the coupling constants in the amplitudes need to be specified.
The coupling constants for the $\psi SH$, $\psi HH$ and $\psi TH$ vertices in Eqs.~(\ref{eq:psiDD})--(\ref{eq:psiD_2D*}) are estimated using the modified Godfrey-Isgur (MGI) model together with the quark-pair-creation (QPC) model~\cite{Liu:2024tgq}. In our calculation, $\psi(4660)$ is assumed to be a pure $\psi(5S)$ state~\cite{Ding:2007rg}. By matching the decay amplitudes derived from the QPC model to the covariant amplitudes in Eqs.~(\ref{eq:psiDD})--(\ref{eq:psiD_2D*}), we determine the effective couplings, as summarized in Table~\ref{tab:coupling}. Details of this procedure are given in Appendix~\ref{MGI Model}.

Considering the heavy quark limit and chiral symmetry, the coupling constants between charmed mesons and Goldstone bosons are given by 
\begin{eqnarray}
&&g_{D^* D \mathcal{M}} = -\frac{g_{D^* D^* \mathcal{M}}}{\sqrt{m_{D^*} m_D}} = -\frac{2 g}{f_\pi} \sqrt{m_{D^*} m_D} \, ,\\
&&g_{D D_0 \mathcal{M}} = - g_{D^* D_1^{\prime} \mathcal{M}} = \frac{2 h}{f_\pi} \, ,\\
&&g_{D^* D_1 \mathcal{M}} = -2\sqrt{\frac{2}{3}} \frac{h^\prime}{\Lambda_\chi f_\pi} \sqrt{m_{D^*} m_{D_1}} \, ,\\
&&g_{D D_2 \mathcal{M}} = -\frac{g_{D^* D_2 \mathcal{M}}}{\sqrt{m_{D_2} m_D}} = -\frac{4 h^{\prime}}{\Lambda_{\chi} f_\pi} \sqrt{m_{D_2} m_D}.
\end{eqnarray}
Here $g =0.59$ is determined from the partial width of $D^* \to D \pi$, $f_\pi=132$ MeV is the pion decay constant, $\Lambda_\chi \simeq 1$ GeV is the chiral symmetry breaking scale, and $h=0.56$, $h^\prime=0.43$~\cite{Colangelo:2012xi}.

For the coupling constants between the molecular states and their components, we employ the compositeness relation to estimate their values~\cite{Weinberg:1962hj}. For the $Z_c(3900)$, we consider two channels, $J/\psi \pi$ and $D \bar{D}^*$, while for the $Z_c(4020)$, the channels $h_c \pi$ and $D^* \bar{D}^*$ are included. The total compositeness relation for the two-channel case reads~\cite{Meissner:2015mza,Guo:2019kdc}
\begin{align}
    X&= X_1 + X_2 \nonumber\\
    &=  |g_1|^2 \Big|\frac{\partial G_1^{II}(s_R)}{\partial s} \Big| +  |g_2|^2 \Big|\frac{\partial G_2^{II}(s_R)}{\partial s} \Big|,\label{eq:X}\\
    \Gamma_R &= \Gamma_1 + \Gamma_2 \nonumber \\
    & = |g_1|^2 \frac{q_1(M_R^2)}{8 \pi M_R^2}\nonumber\\
    &+ |g_2|^2 \int_{m_{th}}^{M_R + 2 \Gamma_R} dE \frac{q_2 (E^2)}{16 {\pi}^2 E^2} \frac{\Gamma_R}{ (M_R - E)^2 + \frac{\Gamma_R^2}{4}},\nonumber\\
    \label{eq:Gamma}
\end{align}
where $ G_j^{II}(s)$ denotes the one-loop two-point function $G(s)$ on the second Riemann sheet~\cite{Oller:2000fj,Oller:2006jw}. The quantity $g_j$ ($j=1,2$) is the effective coupling between the resonance and the corresponding two-particle channel. The lighter channel is labeled by $1$, and the heavier channel by $2$. In this work, we follow the method proposed in Refs.~\cite{Guo:2020vmu,Kang:2016ezb} to estimate these couplings. In terms of the scattering length $a$ and the effective range $r$, the compositeness $X$ is expressed as
\begin{eqnarray}
    X &=& \left( \frac{2 r}{a}-1 \right)^{-1/2}.
\end{eqnarray}
For a resonance pole at $E_R=M_R-i\Gamma_R/2$, the scattering length and effective range are determined by
\begin{eqnarray}
   && a=-\frac{2k_i}{k_r^2+k_i^2},\ \ r=-\frac{1}{k_i}, \\
   && k_r+ik_i\equiv \left[\frac{2m_1 m_2}{m_1+m_2}(E_R-m_1-m_2)\right]^{-1/2}\ ,
\end{eqnarray}
where $m_1$ and $m_2$ are the masses of the scattering particles. By solving Eqs.~(\ref{eq:X}) and (\ref{eq:Gamma}), we obtain
\begin{eqnarray}
    g_{Z_c} = 8.27~\mathrm{GeV}, \qquad g_{Z_c^{\prime}} = 3.29.
\end{eqnarray}
The coupling $g_{Z_{cs}}$ in Eq.~(\ref{eq:Zcs}) is estimated using the same formalism as that for $Z_c(3900)$, except that in the present case the coupled channels are taken to be $J/\psi K$ and $D_s^*D$ ($D^* D_s$).

For the effective coupling constant between the $Z_2(4250)$ and its components, since this state lies slightly below the $D_1 \bar{D}$ threshold, the effective coupling is related to the probability of finding the $D_1 \bar{D}$ component in the physical wave function and to the binding energy, $\epsilon_{Z_2} = m_D + m_{D_1} - m_{Z_2}$, through
\begin{align}
    g_{Z_2}^2 = 16 \pi (m_D + m_{D_1})^2 c^2\sqrt{\frac{2 \epsilon_{Z_2}}{\mu}}.
\end{align}
Here $\mu = m_D m_{D_1}/(m_D + m_{D_1})$ is the reduced mass. For a pure bound state, $c^2=1$.

\section{Numerical results}
\label{sec:numerical}

In this section, we present the numerical results for the processes considered in this work and discuss the main dynamical features of the corresponding triangle-loop mechanisms. The relevant masses are taken as the central values from the PDG~\cite{ParticleDataGroup:2024cfk}. Unless otherwise specified, the branching fractions quoted below refer to the charged modes displayed in the figure captions. Since the cutoff parameter $\Lambda$ in the monopole form factor is not fixed, we vary it in the range $\Lambda=3$--$6$ GeV to estimate the model dependence associated with the off-shell behavior of the exchanged charmed meson. This range is sufficiently above the masses of the exchanged particles and therefore allows us to test whether the predicted rates are mainly driven by the near-threshold loop dynamics or by the ultraviolet tail of the loop integral.

\subsection{$\psi(4660) \to Z_c(3900) \pi$ and $\psi(4660) \to Z_c(4020) \pi$}

\begin{figure}
	\includegraphics[width=0.85\hsize]{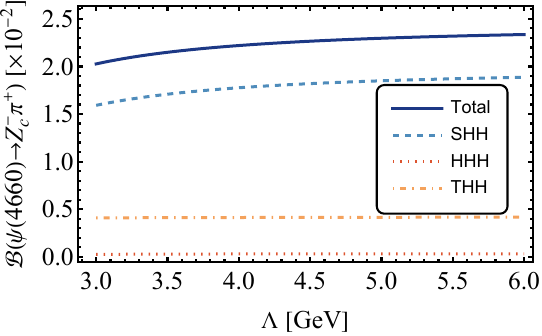}
	\caption{$\Lambda$ dependence of the branching ratio for $\psi(4660) \to Z_c(3900)^- \pi^+$ through the rescattering processes in Fig.~\ref{fig:triangle Zc}. }
	\label{fig:BR Zc(3900)}
\end{figure}

The $\Lambda$ dependence of the branching ratio for $\psi(4660) \to Z_c(3900) \pi$ is shown in Fig.~\ref{fig:BR Zc(3900)}. As $\Lambda$ increases, the monopole form factor suppresses the high-momentum part of the loop integral less strongly, and hence the branching ratio increases slightly. Quantitatively, however, the variation remains modest:
\begin{align}
    \mathcal{B}[\psi(4660) \to Z_c(3900) \pi] \simeq (2.03\text{--}2.34) \times 10^{-2}\,.
\end{align}
The change over the full cutoff interval is only at the level of about $15\%$, indicating that the result is not dominated by short-distance contributions parametrized by the form factor. Instead, the loop amplitude is mainly controlled by the kinematic region in which the intermediate charmed mesons can approach their mass shells. This behavior is consistent with the molecular interpretation of $Z_c(3900)$, for which the coupling to the nearby $D\bar D^*$ threshold plays a central role.

All individual loop contributions increase with $\Lambda$, but their relative hierarchy is rather stable. The $SHH$ channel [Figs.~\ref{fig:triangle Zc}(g) and~\ref{fig:triangle Zc}(h)] is about one order of magnitude larger than the $THH$ and $HHH$ contributions. The origin of this enhancement is twofold. First, the transitions $\psi(4660) \to D_0 \bar{D}^*$ and $\psi(4660) \to D_1^{\prime} \bar D$ proceed through an $S$ wave, while the $\psi(4660) \to D^{(*)} \bar{D}^{(*)}$ and $\psi(4660) \to D_1 \bar D/D_2 \bar D^{(*)}$ transitions occur through $P$- and $D$-wave couplings, respectively. The latter channels therefore suffer from centrifugal suppression. Second, the effective $\psi SH$ couplings listed in Table~\ref{tab:coupling} are numerically favorable for the $D_0\bar D^*$ and $D_1^\prime\bar D$ intermediate states. As a consequence, the $SHH$ loops provide the leading mechanism for producing $Z_c(3900)$ in $\psi(4660)$ decays. The predicted branching fraction at the percent level suggests that this channel should be experimentally accessible if the $\psi(4660)$ has a sizable $\psi(5S)$ component and if the $Z_c(3900)$ couples strongly to $D\bar D^*$.

\begin{figure}
	\includegraphics[width=0.85\hsize]{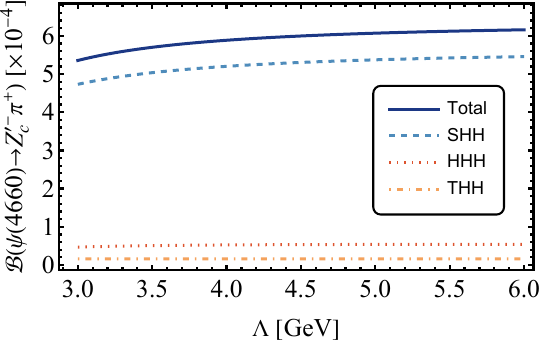}
	\caption{$\Lambda$ dependence of the branching ratio for $\psi(4660) \to Z_c(4020)^- \pi^+$ through the rescattering processes in Fig.~\ref{fig:triangle Zcprime}. }
	\label{fig:BR Zc(4020)}
\end{figure}

Figure~\ref{fig:BR Zc(4020)} shows the corresponding result for $\psi(4660) \to Z_c(4020) \pi$. The cutoff dependence follows the same qualitative trend as in the $Z_c(3900)\pi$ channel, namely a slow increase with $\Lambda$:
\begin{align}
    \mathcal{B}[\psi(4660) \to Z_c(4020) \pi] \simeq (5.36\text{--}6.16) \times 10^{-4}.
\end{align}
The mild cutoff sensitivity again implies that the predicted rate is relatively stable against reasonable variations of the form factor. Similar to $\psi(4660) \to Z_c(3900) \pi$, the $SHH$ loop remains the dominant contribution and exceeds the $THH$ and $HHH$ terms by roughly one order of magnitude. A notable difference is that in the $Z_c(4020)\pi$ channel the $HHH$ contribution is larger than the $THH$ one. 

The total production rate of $Z_c(4020)\pi$ is nevertheless more than one order of magnitude smaller than that of $Z_c(3900)\pi$. Besides the somewhat smaller phase space, this suppression mainly reflects the smaller effective coupling of $Z_c(4020)$ to its $D^*\bar D^*$ component compared with the $Z_c(3900)D\bar D^*$ coupling obtained from the compositeness relation. The different spin structures in the $Z_c^\prime D^*\bar D^*$ vertex also introduce additional momentum dependence. Therefore, the relative size of the two branching fractions provides a useful test of the molecular assignments of $Z_c(3900)$ and $Z_c(4020)$ within the same production framework.

\subsection{$\psi(4660) \to Z_{cs}(3985) K$}

\begin{figure}
	\includegraphics[width=0.85\hsize]{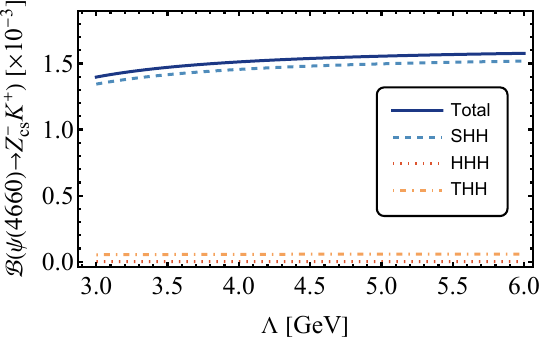}
	\caption{$\Lambda$ dependence of the branching ratio for $\psi(4660) \to Z_{cs}^- K^+$ through the rescattering processes. }
	\label{fig:BR Zcs}
\end{figure}

The process $\psi(4660) \to Z_{cs}(3985) K$ is studied through the triangle diagrams shown in Fig.~\ref{fig:triangle Zcs}, with the coupling constants presented in Table~\ref{tab:coupling}. The branching ratio as a function of $\Lambda$ is displayed in Fig.~\ref{fig:BR Zcs}. Similar to the $Z_c(3900)\pi$ case, it increases only mildly when $\Lambda$ is varied from $3$ to $6$ GeV. 
The resulting branching fraction is
\begin{align}
    \mathcal{B}[\psi(4660) \to Z_{cs}(3985) K] \simeq (1.40\text{--} 1.58) \times 10^{-3}\,.
\end{align}
This value is about one order of magnitude smaller than that for $\psi(4660)\to Z_c(3900)\pi$, but it is still larger than the predicted rate for $\psi(4660)\to Z_c(4020)\pi$. The suppression relative to $Z_c(3900)\pi$ reflects the combined effects of SU(3)-flavor breaking, the reduced phase space caused by the kaon mass, and the different charmed-strange intermediate thresholds. In particular, the heavier strange mesons shift the loop kinematics away from the most favorable near-on-shell region, while the kaon emission vertex also carries different momentum dependence from pion emission.

The hierarchy among the $SHH$, $THH$, and $HHH$ contributions is the same as that in $\psi(4660)\to Z_c(3900)\pi$: the $SHH$ loop dominates, whereas the $THH$ and $HHH$ contributions are subleading, with the $THH$ term larger than the $HHH$ one. This pattern again emphasizes the importance of the $S$-wave coupling of $\psi(4660)$ to the positive-parity charmed mesons in the $S$ doublet. The predicted branching fraction of order $10^{-3}$ indicates that $\psi(4660)\to Z_{cs}K$ can serve as a useful channel for testing the strange partner assignment of $Z_{cs}(3985)$ and for quantifying SU(3)-breaking effects in hidden-charm molecular production.

\subsection{$\psi(4660) \to Z_2(4250) \pi$}

\begin{figure}
	\includegraphics[width=0.85\hsize]{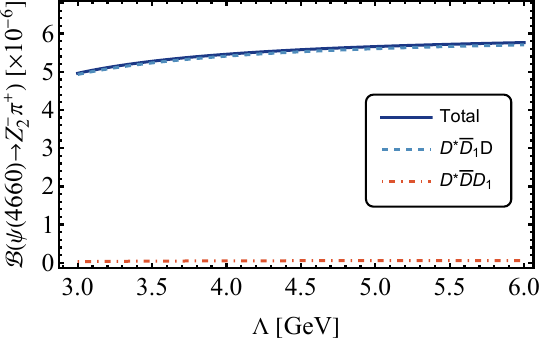}
	\caption{$\Lambda$ dependence of the branching ratio for $\psi(4660) \to Z_2(4250)^- \pi^+$ through the rescattering processes. }
	\label{fig:BR Z2}
\end{figure}

The $\Lambda$ dependence of the branching ratio for $\psi(4660) \to Z_2(4250) \pi$ is illustrated in Fig.~\ref{fig:BR Z2}. In this case, the two triangle diagrams with intermediate $D^*\bar{D} D_1$ and $D^*\bar{D}_1 D$ states, shown in Fig.~\ref{fig:triangle Z2}, are included explicitly. The updated numerical result shows a clear hierarchy between these two loop topologies: the contribution from the $D^*\bar{D}_1 D$ loop is much larger than that from the $D^*\bar{D} D_1$ loop over the whole cutoff range. As a result, the total branching fraction is essentially saturated by the $D^*\bar{D}_1 D$ contribution, while the $D^*\bar{D} D_1$ loop only gives a small correction.

This hierarchy indicates that the production of $Z_2(4250)$ is rather sensitive to the detailed arrangement of the intermediate charmed mesons in the triangle loop. Since $Z_2(4250)$ is assumed to be dominated by the $D_1\bar D$ ($\bar{D}_1 D$) molecular component, the $D^*\bar{D}_1 D$ topology provides a more favorable kinematic matching to the final molecular vertex. In this topology, the intermediate mesons participating in the formation of $Z_2(4250)$ can stay closer to their mass shells, and the loop amplitude is therefore less suppressed by off-shell effects. By contrast, the $D^*\bar{D} D_1$ topology has a less favorable virtuality distribution and is more strongly suppressed in the same cutoff region.

Unlike the previous channels, the total result is almost insensitive to the cutoff parameter in the range $\Lambda=3$--$6$ GeV. This weak dependence suggests that the dominant $D^*\bar{D}_1 D$ loop is mostly restricted by the available phase space and by the heavy intermediate thresholds rather than by the ultraviolet part of the form factor. Numerically, we obtain
\begin{equation}
    \mathcal{B}[\psi(4660) \to Z_2(4250) \pi] = (5.54\text{--}6.35) \times 10^{-6} .
\end{equation}
Compared with the $Z_c(3900)\pi$ and $Z_c(4020)\pi$ modes, the production rate for $Z_2(4250)\pi$ is significantly smaller. Several factors contribute to this suppression.
The final state lies closer to the upper end of the available phase space in $\psi(4660)$ decay. Moreover, the coupling of $\psi(4660)$ to the $D^* \bar{D}_1$ channel is constrained by heavy-quark spin symmetry to proceed through a relative $D$ wave, which leads to an additional centrifugal-barrier suppression.
In addition, the $Z_2D_1\bar D$ coupling is tied to the binding energy and to the probability of the $D_1\bar D$ component in the physical wave function; if the molecular component is smaller than the pure bound-state limit adopted here, the branching fraction would be further reduced.
Thus, observation of this channel would be challenging, but the pronounced dominance of the $D^*\bar{D}_1 D$ loop provides a useful dynamical feature that can be tested once more precise data or upper limits become available.

Combining the four channels, we find a hierarchy
\begin{equation}
\begin{aligned}
& \mathcal{B}[\psi(4660) \to Z_c(3900)\pi] 
 > \mathcal{B}[\psi(4660) \to Z_{cs}(3985)K] \\
&> \mathcal{B}[\psi(4660) \to Z_c(4020)\pi] 
> \mathcal{B}[\psi(4660) \to Z_2(4250)\pi] .
\end{aligned}
\end{equation}
This ordering reflects the interplay of molecular couplings, available phase space, heavy-meson partial waves, and SU(3)-flavor breaking. The dominance of $SHH$ loops in the first three channels is a robust consequence of the $S$-wave $\psi SH$ transition, while the strong suppression of $Z_2(4250)\pi$ highlights the sensitivity of higher-mass molecular candidates to the detailed dynamics of excited charmed mesons. These features make $\psi(4660)$ decays a useful laboratory for comparing different molecular configurations within a unified loop mechanism.

\section{SUMMARY}
\label{sec:summary}

In this work, we have studied the production of hidden-charm molecular candidates in $\psi(4660)$ decays through intermediate charmed-meson triangle loops. Treating $\psi(4660)$ as a conventional $\psi(5S)$ charmonium state, we evaluated the processes $\psi(4660)\to Z_c(3900)\pi$, $\psi(4660)\to Z_c(4020)\pi$, $\psi(4660)\to Z_{cs}(3985)K$, and $\psi(4660)\to Z_2(4250)\pi$ in a unified framework. The exotic states $Z_c(3900)$, $Z_c(4020)$, $Z_{cs}(3985)$, and $Z_2(4250)$ were assumed to be dominated by the $D\bar D^*$, $D^*\bar D^*$, $D_s\bar D^*/D_s^*\bar D$, and $D_1\bar D$ molecular components, respectively. The effective couplings of $\psi(4660)$ to charmed-meson pairs were estimated with the MGI model combined with the QPC model, while the couplings of the molecular states to their constituents were determined from the compositeness relation.

Our numerical results show that the branching fractions for $\psi(4660) \to Z_c(3900)^- \pi^+$, $\psi(4660) \to Z_c(4020)^- \pi^+$, $\psi(4660) \to Z_{cs}(3985)^- K^+$, and $\psi(4660) \to Z_2(4250)^-\pi^+$ are of the order of $10^{-2}$, $10^{-4}$, $10^{-3}$, and $10^{-6}$, respectively. These rates exhibit only a mild dependence on the cutoff parameter in the monopole form factor, indicating that the main conclusions are not dominated by the short-distance part of the loop integral. Among the four channels, $\psi(4660)\to Z_c(3900)\pi$ is predicted to have the largest branching fraction, while $\psi(4660)\to Z_2(4250)\pi$ is strongly suppressed.

A channel-by-channel analysis further shows that the $SHH$ intermediate loops generally provide the leading contributions in the production of $Z_c(3900)$, $Z_c(4020)$, and $Z_{cs}(3985)$. This dominance can be understood from the $S$-wave coupling of $\psi(4660)$ to the positive-parity $S$-doublet charmed mesons, in contrast to the centrifugal suppression associated with the $P$- and $D$-wave $HH$ and $TH$ transitions. The hierarchy of the predicted rates therefore reflects the combined effects of molecular couplings, available phase space, partial-wave structures, and SU(3)-flavor breaking.

The sizable branching fractions predicted for the $Z_c(3900)\pi$, $Z_{cs}(3985)K$, and, to a lesser extent, $Z_c(4020)\pi$ channels suggest that $\psi(4660)$ decays provide a useful platform for testing the molecular interpretation of charged hidden-charm exotic states. In particular, future measurements of these modes, as well as improved upper limits on the suppressed $Z_2(4250)\pi$ channel, would help clarify the role of charmed-meson loops and the internal structure of these exotic candidates.

\begin{acknowledgments}
\label{sec:acknowledgements}
This work is supported by the National Natural Science Foundation of China under Grants No.~12235018, No.~12475081, and No.~11975165; the Natural Science Foundation of Shandong Province under Grant No.~ZR2025MS04; and the Taishan Scholar Project of Shandong Province.
\end{acknowledgments}

\onecolumngrid
\appendix

\section{The MGI model}
\label{MGI Model}
The interaction between a quark ($q$) and an antiquark ($\bar{q}$) is described by the Hamiltonian 
\begin{equation}
\tilde{H} = (p_1^2 + m_1^2)^{1/2} + (p_2^2 + m_2^2)^{1/2} +\tilde{V}_{eff}( \mathbf{p}, \mathbf{r} ),
\end{equation}
where $m_i$ denotes the mass of the quark $q$ or antiquark $\bar{q}$. The effective potential $\tilde{V}_{eff}(\mathbf{p}, \mathbf{r})$ describes the interaction between $q$ and $\bar{q}$, including both a short-range one-gluon-exchange term $\gamma_{\mu} \otimes \gamma^{\mu}$ and a long-range confinement term $1 \otimes 1$. It is expressed as 
\begin{equation}
    \tilde{V}_{eff}(\mathbf{p}, \mathbf{r}) = H^{conf} + H^{hyp} + H^{so},
    \label{eq:Veff}
\end{equation}
where
\begin{eqnarray}
    H^{conf}  = br - \frac{4 \alpha_s(r)}{3r} + c
\end{eqnarray}
is the spin-independent potential containing a constant term, the linear confining potential $S(r)$, and a one-gluon exchange potential $G(r)$. $\alpha_s(r)$ is the running,
\begin{eqnarray}
    \alpha_s(r) = \sum\limits_k \frac{2 \alpha_k}{\sqrt{\pi}} \int_0^{\gamma_k r} e^{- x^2} dx, 
\end{eqnarray}
where the sum runs from $k=1$ to $3$, $\alpha_k = (0.25, 0.15, 0.2)$, and $\gamma_k = (1/2, \sqrt{10}/2, \sqrt{1000}/2)$.

The color-hyperfine interaction $H^{hyp}$ in Eq.~\eqref{eq:Veff} consists of the spin-spin and tensor terms, given by
\begin{eqnarray}
    H^{hyp} &=& -\frac{\alpha_s(r)}{m_1 m_2} \left[ \frac{8 \pi}{3} \mathbf{S}_1 \cdot \mathbf{S}_2 \delta^3(r) + \frac{1}{r^3} \left( \frac{3 \mathbf{S}_1 \cdot \mathbf{r} \mathbf{S}_2 \cdot \mathbf{r}}{r^2}  \mathbf{S}_1 \mathbf{S}_2\right) \right] \mathbf{F}_1 \cdot \mathbf{F}_2,
    \label{eq:hyp}
\end{eqnarray}
%where $\mathbf{S}_1/\mathbf{S}_2$ denotes the spin of the quark/antiquark. 
where $\mathbf{S}_1$ and $\mathbf{S}_2$ denote the spins of the quark and antiquark, respectively.
$\mathbf{F}$ represents the Gell-Mann color matrix and $\left< \mathbf{F}_1 \cdot \mathbf{F}_2 \right> = -4/3$.

The last term in Eq.~\eqref{eq:Veff} is the spin-orbit interaction,
\begin{eqnarray}
    H^{so} &=& H^{so(cm)} + H^{so(tp)},
\end{eqnarray}
where $H^{so(cm)}$ is the color-magnetic term and $H^{so(tp)}$ is the Thomas-precession term, which can be written as
\begin{eqnarray}
    H^{so(cm)} &=& \frac{\alpha_s(r)}{r^3} \left( \frac{1}{m_1} + \frac{1}{m_2} \right) \left( \frac{\mathbf{S}_1}{m_1} + \frac{\mathbf{S}_2}{m_2} \right) \cdot \mathbf{L} ( \mathbf{F}_1 \cdot \mathbf{F}_2),\\
    H^{so(tp)} &=& \frac{1}{2 r} \frac{\partial H^{conf}}{\partial r} \left( \frac{\mathbf{S}_1}{m_1^2} + \frac{\mathbf{S}_2}{m_2^2} \right) \cdot \mathbf{L},
    \label{eq:sotp}
\end{eqnarray}
where $\mathbf{L}$ is the relative orbital angular momentum between the quark and the antiquark.

The GI model constructed by Godfrey and Isgur is a relativized quark model, where relativistic effects are embedded in the model in two main ways.
First, a smearing function $\rho_{12}(\mathbf{r}-\mathbf{r^{\prime}})$ is introduced to incorporate the effects of internal motion inside a hadron and the nonlocality of interactions between the quark and antiquark. Its general form is 
\begin{eqnarray}
    \tilde{V}(r) &=& \int d^3 \mathbf{r^{\prime}} \rho (\mathbf{r}- \mathbf{r^{\prime}}) V(r^{\prime}),
\end{eqnarray}
where 
\begin{eqnarray}
    \rho_{ij} (\mathbf{r}- \mathbf{r^{\prime}}) &=& \frac{\sigma_{ij}^3}{\pi^{3/2}} e^{- \sigma_{ij}^2 (\mathbf{r}- \mathbf{r^{\prime}})^2 }.
\end{eqnarray}
Here $\sigma_{ij}$ is the smearing parameter, 
\begin{eqnarray}
    \sigma_{ij}^2 = \sigma_0^2 \left[ \frac{1}{2} + \frac{1}{2} \left( \frac{4 m_i m_j}{ (m_i + m_j)^2}\right)^4 \right] + s^2 \left( \frac{2 m_i m_j}{m_i + m_j} \right)^2,
\end{eqnarray}
where $\sigma_0 = 1.791$ GeV is a universal parameter and $s=0.711$. The values of these two parameters, listed in Table~\ref{tab:MGI}, are taken from Ref.~\cite{Wang:2019mhs}.

\begin{table}
\renewcommand\arraystretch{1.5}
\caption{The parameters of the MGI model used in this work.}
\label{tab:MGI}
\begin{ruledtabular}
\begin{tabular}{cccc}
Parameters	& Values&Parameters	& Values\\
\colrule
$m_{u(d)} (\mathrm{GeV})$&$0.22$&$m_s(\mathrm{GeV})$&$0.419$\\
$m_c(\mathrm{GeV})$&$1.65$&$\epsilon_t$& $0.012$\\
$b(\mathrm{GeV}^2)$&$0.2687$&$\epsilon_{so(v)}$&$-0.053$\\
$c(\mathrm{GeV})$& $-0.3673$&$\epsilon_{so(s)}$&$0.083$\\
$\mu(\mathrm{GeV})$&$0.15$&$\epsilon_c$&$-0.084$\\
\end{tabular}
\end{ruledtabular}
\end{table} 

Furthermore, to account for the unquenched effect, it is common to replace the linear potential with the screening potential 
\begin{eqnarray}
    br \to \frac{b(1-e^{-\mu r})}{\mu},
\end{eqnarray}
where $\mu$ represents the strength of the screening effect. Therefore, the spin-independent Coulomb term $\tilde{G}(r)$ in the MGI model becomes
\begin{eqnarray}
    \tilde{G}_{ij}(r) = \int d^3 \mathbf{r^{\prime}} \rho (\mathbf{r}- \mathbf{r^{\prime}}) G(r^{\prime})= - \sum\limits_{k=1}\limits^3 \frac{4 \alpha_k}{3r}erf(\tau_{kij}r).
\end{eqnarray}
Here the $\tau_{kij}$ are obtained from 
\begin{eqnarray}
    \tau_{kij}= \frac{1}{\sqrt{\frac{1}{\sigma_{ij}^2} + \frac{1}{\gamma_k^2}}}.
\end{eqnarray}
The screened confinement term $\tilde{S}_{12}(r)$ is expressed as
\begin{align}
    \tilde{S}_{12}(r) &= \int d^3 \mathbf{r^{\prime}} \rho (\mathbf{r}- \mathbf{r^{\prime}}) S(r^{\prime}) \nonumber \\
    &= \frac{b}{\mu r} \Bigg[ r + e^{\frac{\mu^2}{4 \sigma^2}+ \mu r} \frac{\mu + 2r\sigma^2}{2 \sigma^2} \Bigg( \frac{1}{\sqrt{\pi}} \int_0^{\frac{\mu + 2 r \sigma^2}{2 \sigma}} e^{-x^2} dx - \frac{1}{2} \Bigg) \nonumber\\
    &- e^{\frac{\mu^2}{4 \sigma^2}- \mu r} \frac{\mu - 2r\sigma^2}{2 \sigma^2} \Bigg( \frac{1}{\sqrt{\pi}} \int_0^{\frac{\mu - 2 r \sigma^2}{2 \sigma}} e^{-x^2} dx - \frac{1}{2} \Bigg) \Bigg] +c.
\end{align}

Second, a general formula of the potential should depend on the center-of-mass (CM) momentum of the interacting quarks. This effect is taken into account by introducing momentum-dependent factors in the interactions and these factors become unity in the nonrelativistic limit. In a semiquantitatively relativistic treatment, the smeared Coulomb term $\tilde{G}(r)$ and the smeared hyperfine interactions $\tilde{V}^i$ are modified according to
\begin{eqnarray}
    \tilde{G}(r) &=& \left( 1 + \frac{p^2}{E_1 E_2} \right)^{1/2} \tilde{G}(r) \left( 1 + \frac{p^2}{E_1 E_2} \right)^{1/2},\\
    \tilde{V}^i(r) &=& \left( \frac{m_1 m_2}{E_1 E_2} \right)^{1/2 + \epsilon_i} \tilde{V}^i(r) \left( \frac{m_1 m_2}{E_1 E_2} \right)^{1/2 + \epsilon_i}.
    \label{eq:Vi}
\end{eqnarray}
Here $E_1$ and $E_2$ are the energies of the two constituent quarks. $\tilde{V}^i(r)$ denotes the contact term,  the tensor term, the vector spin-orbit, and the scalar spin-orbit terms of Eqs.~(\ref{eq:hyp})-(\ref{eq:sotp}). The parameters $\epsilon_i= \epsilon_c, \epsilon_t, \epsilon_{so(v)}$, and $\epsilon_{so(s)}$ represent the relativistic corrections to $\tilde{V}^{cont}$, $\tilde{V}^{tens}$, $\tilde{V}^{so(v)}$, and $\tilde{V}^{so(s)}$, respectively. 

The Hamiltonian can be written as
\begin{eqnarray}
    H = (p^2 + m_1^2)^{\frac{1}{2}} + (p^2 + m_2^2)^{\frac{1}{2}} + \tilde{G}_{12} + \tilde{S}_{12}(r) + \tilde{V}^{cont} + \tilde{V}^{so(v)} +\tilde{V}^{so(s)}.
    \label{eq:H}
\end{eqnarray}
The explicit forms of these spin-dependent terms in Eq.~\eqref{eq:Vi} are given by 
\begin{eqnarray}
    \tilde{V}^{cont} &=& \frac{2 \mathbf{S}_1 \cdot \mathbf{S}_2}{3 m_1m_2} \nabla^2 \tilde{G}^c_{12}, \\
     \tilde{V}^{tens} &=& -\left( \frac{3 \mathbf{S}_1 \cdot \mathbf{r} \mathbf{S}_2 \cdot \mathbf{r}/r^2 - \mathbf{S}_1 \cdot \mathbf{S}_2}{3m_1 m_2} \right) \left( \frac{\partial^2}{\partial r^2} - \frac{1}{r} \frac{\partial}{\partial r} \right) \tilde{G}_{12}^t, \\
     \tilde{V}^{so(v)} &=& \frac{\mathbf{S}_1 \cdot \mathbf{L}}{2 m_1^2} \frac{1}{r} \frac{\partial \tilde{G}^{so(v)}_{11}}{\partial r} + \frac{\mathbf{S}_2 \cdot \mathbf{L}}{2 m_2^2} \frac{1}{r} \frac{\partial \tilde{G}^{so(v)}_{22}}{\partial r} + \frac{(\mathbf{S}_1 + \mathbf{S}_2) \cdot \mathbf{L}}{m_1 m_2} \frac{1}{r} \frac{\partial \tilde{G}_{12}^{so(v)}}{\partial r}, \\
     \tilde{V}^{so(s)} &=& - \frac{\mathbf{S}_1 \cdot \mathbf{L}}{2 m_1^2} \frac{1}{r} \frac{\partial \tilde{S}_{11}^{so(s)}}{\partial r} - \frac{\mathbf{S}_2 \cdot \mathbf{L}}{2 m_2^2} \frac{1}{r} \frac{\partial \tilde{S}_{22}^{so(s)}}{\partial r}.
\end{eqnarray}
By diagonalizing the Hamiltonian matrix in Eq.~\eqref{eq:H} within the simple harmonic oscillator (SHO) basis, the spatial wave functions of the mesons can be obtained. For simplicity, we set $\beta=0.5$.
\twocolumngrid

%-----------------------------------------------------------

\bibliography{manuscript_01.bib}

@article{Brambilla:2019esw,
    author = "Brambilla, Nora and Eidelman, Simon and Hanhart, Christoph and Nefediev, Alexey and Shen, Cheng-Ping and Thomas, Christopher E. and Vairo, Antonio and Yuan, Chang-Zheng",
    title = "{The $XYZ$ states: experimental and theoretical status and perspectives}",
    eprint = "1907.07583",
    archivePrefix = "arXiv",
    primaryClass = "hep-ex",
    reportNumber = "TUM-EFT 125/19",
    doi = "10.1016/j.physrep.2020.05.001",
    journal = "Phys. Rept.",
    volume = "873",
    pages = "1--154",
    year = "2020"
}

@article{Chen:2016qju,
    author = "Chen, Hua-Xing and Chen, Wei and Liu, Xiang and Zhu, Shi-Lin",
    title = "{The hidden-charm pentaquark and tetraquark states}",
    eprint = "1601.02092",
    archivePrefix = "arXiv",
    primaryClass = "hep-ph",
    doi = "10.1016/j.physrep.2016.05.004",
    journal = "Phys. Rept.",
    volume = "639",
    pages = "1--121",
    year = "2016"
}

@article{Guo:2017jvc,
    author = "Guo, Feng-Kun and Hanhart, Christoph and Mei{\ss}ner, Ulf-G. and Wang, Qian and Zhao, Qiang and Zou, Bing-Song",
    title = "{Hadronic molecules}",
    eprint = "1705.00141",
    archivePrefix = "arXiv",
    primaryClass = "hep-ph",
    doi = "10.1103/RevModPhys.90.015004",
    journal = "Rev. Mod. Phys.",
    volume = "90",
    number = "1",
    pages = "015004",
    year = "2018",
    note = "[Erratum: Rev.Mod.Phys. 94, 029901 (2022)]"
}

@article{Liu:2019zoy,
    author = "Liu, Yan-Rui and Chen, Hua-Xing and Chen, Wei and Liu, Xiang and Zhu, Shi-Lin",
    title = "{Pentaquark and Tetraquark states}",
    eprint = "1903.11976",
    archivePrefix = "arXiv",
    primaryClass = "hep-ph",
    doi = "10.1016/j.ppnp.2019.04.003",
    journal = "Prog. Part. Nucl. Phys.",
    volume = "107",
    pages = "237--320",
    year = "2019"
}

@article{Meng:2022ozq,
    author = "Meng, Lu and Wang, Bo and Wang, Guang-Juan and Zhu, Shi-Lin",
    title = "{Chiral perturbation theory for heavy hadrons and chiral effective field theory for heavy hadronic molecules}",
    eprint = "2204.08716",
    archivePrefix = "arXiv",
    primaryClass = "hep-ph",
    doi = "10.1016/j.physrep.2023.04.003",
    journal = "Phys. Rept.",
    volume = "1019",
    pages = "1--149",
    year = "2023"
}

@misc{Bai:2026atm,
    author = "Bai, Zi-Yue and Chen, Dian-Yong and Qi-Huang and Liu, Xiang and Luo, Si-Qiang and Wang, Jun-Zhang",
    title = "{Unquenched Charmonium and Beyond}",
    eprint = "2602.19887",
    archivePrefix = "arXiv",
    primaryClass = "hep-ph",
    month = "2",
    year = "2026"
}

@article{BESIII:2013ris,
    author = "Ablikim, M. and others",
    collaboration = "BESIII",
    title = "{Observation of a Charged Charmoniumlike Structure in $e^+e^- \to \pi^+\pi^- J/\psi$ at $\sqrt{s}$ = 4.26 GeV}",
    eprint = "1303.5949",
    archivePrefix = "arXiv",
    primaryClass = "hep-ex",
    doi = "10.1103/PhysRevLett.110.252001",
    journal = "Phys. Rev. Lett.",
    volume = "110",
    pages = "252001",
    year = "2013"
}

@article{Belle:2013yex,
    author = "Liu, Z. Q. and others",
    collaboration = "Belle",
    title = "{Study of $e^+e^- \to \pi^+ \pi^- J/\psi$ and Observation of a Charged Charmoniumlike State at Belle}",
    eprint = "1304.0121",
    archivePrefix = "arXiv",
    primaryClass = "hep-ex",
    reportNumber = "BELLE-PREPRINT-2013-6, KEK-PREPRINT-2013-2",
    doi = "10.1103/PhysRevLett.110.252002",
    journal = "Phys. Rev. Lett.",
    volume = "110",
    pages = "252002",
    year = "2013",
    note = "[Erratum: Phys.Rev.Lett. 111, 019901 (2013)]"
}

@article{BESIII:2013qmu,
    author = "Ablikim, M. and others",
    collaboration = "BESIII",
    title = "{Observation of a charged $(D\bar{D}^{*})^\pm$ mass peak in $e^{+}e^{-} \to \pi D\bar{D}^{*}$ at $\sqrt{s} =$ 4.26 GeV}",
    eprint = "1310.1163",
    archivePrefix = "arXiv",
    primaryClass = "hep-ex",
    doi = "10.1103/PhysRevLett.112.022001",
    journal = "Phys. Rev. Lett.",
    volume = "112",
    number = "2",
    pages = "022001",
    year = "2014"
}

@article{BESIII:2015cld,
    author = "Ablikim, M. and others",
    collaboration = "BESIII",
    title = "{Observation of $Z_c(3900)^{0}$ in $e^+e^-\to\pi^0\pi^0 J/\psi$}",
    eprint = "1506.06018",
    archivePrefix = "arXiv",
    primaryClass = "hep-ex",
    doi = "10.1103/PhysRevLett.115.112003",
    journal = "Phys. Rev. Lett.",
    volume = "115",
    number = "11",
    pages = "112003",
    year = "2015"
}

@article{BESIII:2015ntl,
    author = "Ablikim, M. and others",
    collaboration = "BESIII",
    title = "{Observation of a Neutral Structure near the $D\bar{D}^{*}$ Mass Threshold in $e^{+}e^{-}\to (D \bar{D}^*)^0\pi^0$ at $\sqrt{s}$ = 4.226 and 4.257 GeV}",
    eprint = "1509.05620",
    archivePrefix = "arXiv",
    primaryClass = "hep-ex",
    doi = "10.1103/PhysRevLett.115.222002",
    journal = "Phys. Rev. Lett.",
    volume = "115",
    number = "22",
    pages = "222002",
    year = "2015"
}

@article{BESIII:2013ouc,
    author = "Ablikim, M. and others",
    collaboration = "BESIII",
    title = "{Observation of a Charged Charmoniumlike Structure $Z_c(4020)$ and Search for the $Z_c(3900)$ in $e^+e^- \to \pi^+ \pi^-h_c$}",
    eprint = "1309.1896",
    archivePrefix = "arXiv",
    primaryClass = "hep-ex",
    doi = "10.1103/PhysRevLett.111.242001",
    journal = "Phys. Rev. Lett.",
    volume = "111",
    number = "24",
    pages = "242001",
    year = "2013"
}

@article{BESIII:2013mhi,
    author = "Ablikim, M. and others",
    collaboration = "BESIII",
    title = "{Observation of a charged charmoniumlike structure in $e^+e^- \to (D^{*} \bar{D}^{*})^{\pm} \pi^\mp$ at $\sqrt{s} = 4.26$ GeV}",
    eprint = "1308.2760",
    archivePrefix = "arXiv",
    primaryClass = "hep-ex",
    doi = "10.1103/PhysRevLett.112.132001",
    journal = "Phys. Rev. Lett.",
    volume = "112",
    number = "13",
    pages = "132001",
    year = "2014"
}

@article{BESIII:2014gnk,
    author = "Ablikim, M. and others",
    collaboration = "BESIII",
    title = "{Observation of $e^+e^- \to \pi^0 \pi^0 h_c$ and a Neutral Charmoniumlike Structure $Z_c(4020)^0$}",
    eprint = "1409.6577",
    archivePrefix = "arXiv",
    primaryClass = "hep-ex",
    doi = "10.1103/PhysRevLett.113.212002",
    journal = "Phys. Rev. Lett.",
    volume = "113",
    number = "21",
    pages = "212002",
    year = "2014"
}

@article{BESIII:2015tix,
    author = "Ablikim, M. and others",
    collaboration = "BESIII",
    title = "{Observation of a neutral charmoniumlike state $Z_c(4025)^0$ in $e^{+} e^{-} \to (D^{*} \bar{D}^{*})^{0} \pi^0$}",
    eprint = "1507.02404",
    archivePrefix = "arXiv",
    primaryClass = "hep-ex",
    doi = "10.1103/PhysRevLett.115.182002",
    journal = "Phys. Rev. Lett.",
    volume = "115",
    number = "18",
    pages = "182002",
    year = "2015"
}

@article{Wang:2013cya,
    author = "Wang, Qian and Hanhart, Christoph and Zhao, Qiang",
    title = "{Decoding the riddle of $Y(4260)$ and $Z_c(3900)$}",
    eprint = "1303.6355",
    archivePrefix = "arXiv",
    primaryClass = "hep-ph",
    doi = "10.1103/PhysRevLett.111.132003",
    journal = "Phys. Rev. Lett.",
    volume = "111",
    number = "13",
    pages = "132003",
    year = "2013"
}

@article{Aceti:2014uea,
    author = "Aceti, F. and Bayar, M. and Oset, E. and Martinez Torres, A. and Khemchandani, K. P. and Dias, Jorgivan Morais and Navarra, F. S. and Nielsen, M.",
    title = "{Prediction of an $I=1$ $D \bar D^*$ state and relationship to the claimed $Z_c(3900)$, $Z_c(3885)$}",
    eprint = "1401.8216",
    archivePrefix = "arXiv",
    primaryClass = "hep-ph",
    doi = "10.1103/PhysRevD.90.016003",
    journal = "Phys. Rev. D",
    volume = "90",
    number = "1",
    pages = "016003",
    year = "2014"
}

@article{Guo:2013sya,
    author = "Guo, Feng-Kun and Hidalgo-Duque, Carlos and Nieves, Juan and Valderrama, Manuel Pavon",
    title = "{Consequences of Heavy Quark Symmetries for Hadronic Molecules}",
    eprint = "1303.6608",
    archivePrefix = "arXiv",
    primaryClass = "hep-ph",
    doi = "10.1103/PhysRevD.88.054007",
    journal = "Phys. Rev. D",
    volume = "88",
    pages = "054007",
    year = "2013"
}

@article{Cui:2013yva,
    author = "Cui, Chun-Yu and Liu, Yong-Lu and Chen, Wen-Bo and Huang, Ming-Qiu",
    title = "{Could $Z_{c}(3900)$ be a $I^{G}J^{P}=1^{+}1^{+}$ $D^{*}\bar{D}$ molecular state?}",
    eprint = "1304.1850",
    archivePrefix = "arXiv",
    primaryClass = "hep-ph",
    doi = "10.1088/0954-3899/41/7/075003",
    journal = "J. Phys. G",
    volume = "41",
    pages = "075003",
    year = "2014"
}

@article{Zhang:2013aoa,
    author = "Zhang, Jian-Rong",
    title = "{Improved QCD sum rule study of $Z_{c}(3900)$ as a $\bar{D}D^{*}$ molecular state}",
    eprint = "1304.5748",
    archivePrefix = "arXiv",
    primaryClass = "hep-ph",
    doi = "10.1103/PhysRevD.87.116004",
    journal = "Phys. Rev. D",
    volume = "87",
    number = "11",
    pages = "116004",
    year = "2013"
}

@article{Chen:2013omd,
    author = "Chen, Wei and Steele, T. G. and Du, Meng-Lin and Zhu, Shi-Lin",
    title = "{$D^*\bar D^*$ molecule interpretation of $Z_c(4025)$}",
    eprint = "1308.5060",
    archivePrefix = "arXiv",
    primaryClass = "hep-ph",
    doi = "10.1140/epjc/s10052-014-2773-y",
    journal = "Eur. Phys. J. C",
    volume = "74",
    number = "2",
    pages = "2773",
    year = "2014"
}

@article{Li:2013xia,
    author = "Li, Gang",
    title = "{Hidden-charmonium decays of $Z_c(3900)$ and $Z_c(4025)$ in intermediate meson loops model}",
    eprint = "1304.4458",
    archivePrefix = "arXiv",
    primaryClass = "hep-ph",
    doi = "10.1140/epjc/s10052-013-2621-5",
    journal = "Eur. Phys. J. C",
    volume = "73",
    number = "11",
    pages = "2621",
    year = "2013"
}

@article{Li:2014pfa,
    author = "Li, Gang and Liu, Xiao Hai and Zhou, Zhu",
    title = "{More hidden heavy quarkonium molecules and their discovery decay modes}",
    eprint = "1409.0754",
    archivePrefix = "arXiv",
    primaryClass = "hep-ph",
    doi = "10.1103/PhysRevD.90.054006",
    journal = "Phys. Rev. D",
    volume = "90",
    number = "5",
    pages = "054006",
    year = "2014"
}

@article{Chen:2015igx,
    author = "Chen, Dian-Yong and Dong, Yu-Bing",
    title = "{Radiative decays of the neutral $Z_c(3900)$}",
    eprint = "1510.00829",
    archivePrefix = "arXiv",
    primaryClass = "hep-ph",
    doi = "10.1103/PhysRevD.93.014003",
    journal = "Phys. Rev. D",
    volume = "93",
    number = "1",
    pages = "014003",
    year = "2016"
}

@article{Chen:2016byt,
    author = "Chen, Dian-Yong and Dong, Yu-Bing and Li, Ming-Tao and Wang, Wen-Ling",
    title = "{Pionic transition from $Y(4260)$ to $Z_{c}(3900)$ in a hadronic molecular scenario}",
    doi = "10.1140/epja/i2016-16310-0",
    journal = "Eur. Phys. J. A",
    volume = "52",
    number = "10",
    pages = "310",
    year = "2016"
}

@article{Xiao:2018kfx,
    author = "Xiao, Cheng-Jian and Chen, Dian-Yong and Dong, Yu-Bing and Zuo, Wei and Matsuki, Takayuki",
    title = "{Understanding the $\eta_c\rho$ decay mode of $Z_c^{(\prime)}$ via the triangle loop mechanism}",
    eprint = "1811.04688",
    archivePrefix = "arXiv",
    primaryClass = "hep-ph",
    doi = "10.1103/PhysRevD.99.074003",
    journal = "Phys. Rev. D",
    volume = "99",
    number = "7",
    pages = "074003",
    year = "2019"
}

@article{Braaten:2013boa,
    author = "Braaten, Eric",
    title = "{How the $Z_c(3900)$ Reveals the Spectra of Quarkonium Hybrid and Tetraquark Mesons}",
    eprint = "1305.6905",
    archivePrefix = "arXiv",
    primaryClass = "hep-ph",
    doi = "10.1103/PhysRevLett.111.162003",
    journal = "Phys. Rev. Lett.",
    volume = "111",
    pages = "162003",
    year = "2013"
}

@article{Faccini:2013lda,
    author = "Maiani, L. and Riquer, V. and Faccini, R. and Piccinini, F. and Pilloni, A. and Polosa, A. D.",
    title = "{A $J^{PG}=1^{++}$ Charged Resonance in the $Y(4260) \to \pi^+ \pi^- J/\psi$ Decay?}",
    eprint = "1303.6857",
    archivePrefix = "arXiv",
    primaryClass = "hep-ph",
    doi = "10.1103/PhysRevD.87.111102",
    journal = "Phys. Rev. D",
    volume = "87",
    number = "11",
    pages = "111102",
    year = "2013"
}

@article{Wang:2013llv,
    author = "Wang, Zhi-Gang",
    title = "{Reanalysis of the $Z_c(4020)$, $Z_c(4025)$, $Z(4050)$ and $Z(4250)$ as tetraquark states with QCD sum rules}",
    eprint = "1312.1537",
    archivePrefix = "arXiv",
    primaryClass = "hep-ph",
    doi = "10.1088/0253-6102/63/4/466",
    journal = "Commun. Theor. Phys.",
    volume = "63",
    number = "4",
    pages = "466--480",
    year = "2015"
}

@article{Qiao:2013dda,
    author = "Qiao, Cong-Feng and Tang, Liang",
    title = "{Interpretation of $Z_c(4025)$ as the hidden charm tetraquark states via QCD Sum Rules}",
    eprint = "1308.3439",
    archivePrefix = "arXiv",
    primaryClass = "hep-ph",
    doi = "10.1140/epjc/s10052-014-2810-x",
    journal = "Eur. Phys. J. C",
    volume = "74",
    pages = "2810",
    year = "2014"
}

@article{He:2017lhy,
    author = "He, Jun and Chen, Dian-Yong",
    title = "{$Z_c(3900)/Z_c(3885)$ as a virtual state from $\pi J/\psi-\bar{D}^*D$ interaction}",
    eprint = "1712.05653",
    archivePrefix = "arXiv",
    primaryClass = "hep-ph",
    doi = "10.1140/epjc/s10052-018-5580-z",
    journal = "Eur. Phys. J. C",
    volume = "78",
    number = "2",
    pages = "94",
    year = "2018"
}

@article{Swanson:2014tra,
    author = "Swanson, E. S.",
    title = "{$Z_b$ and $Z_c$ Exotic States as Coupled Channel Cusps}",
    eprint = "1409.3291",
    archivePrefix = "arXiv",
    primaryClass = "hep-ph",
    doi = "10.1103/PhysRevD.91.034009",
    journal = "Phys. Rev. D",
    volume = "91",
    number = "3",
    pages = "034009",
    year = "2015"
}

@article{Szczepaniak:2015eza,
    author = "Szczepaniak, Adam P.",
    title = "{Triangle Singularities and $XYZ$ Quarkonium Peaks}",
    eprint = "1501.01691",
    archivePrefix = "arXiv",
    primaryClass = "hep-ph",
    reportNumber = "JLAB-THY-15-1999",
    doi = "10.1016/j.physletb.2015.06.029",
    journal = "Phys. Lett. B",
    volume = "747",
    pages = "410--416",
    year = "2015"
}

@article{Yu:2024sqv,
    author = "Yu, Kang and Wang, Guang-Juan and Wu, Jia-Jun and Yang, Zhi",
    title = "{Three-coupled-channel analysis of $Z_c(3900)$ involving $D\bar{D}^*$, $\pi J/\psi$, and $\rho\eta$}",
    eprint = "2409.10865",
    archivePrefix = "arXiv",
    primaryClass = "hep-ph",
    doi = "10.1103/PhysRevD.110.114029",
    journal = "Phys. Rev. D",
    volume = "110",
    number = "11",
    pages = "114029",
    year = "2024"
}

@article{Ablikim:2020hsk,
    author = "Ablikim, Medina and others",
    collaboration = "BESIII",
    title = "{Observation of a Near-Threshold Structure in the $K^+$ Recoil-Mass Spectra in $e^+e^- \to K^+(D_s^-D^{*0}+D_s^{*-}D^0$)}",
    eprint = "2011.07855",
    archivePrefix = "arXiv",
    primaryClass = "hep-ex",
    doi = "10.1103/PhysRevLett.126.102001",
    journal = "Phys. Rev. Lett.",
    volume = "126",
    number = "10",
    pages = "102001",
    year = "2021"
}

@article{Aaij:2021ivw,
    author = "Aaij, Roel and others",
    collaboration = "LHCb",
    title = "{Observation of New Resonances Decaying to $J/\psi K^+$ and $J/\psi \phi$}",
    eprint = "2103.01803",
    archivePrefix = "arXiv",
    primaryClass = "hep-ex",
    reportNumber = "LHCb-PAPER-2020-044, CERN-EP-2021-025",
    doi = "10.1103/PhysRevLett.127.082001",
    journal = "Phys. Rev. Lett.",
    volume = "127",
    number = "8",
    pages = "082001",
    year = "2021"
}

@article{Chen:2022asf,
    author = "Chen, Hua-Xing and Chen, Wei and Liu, Xiang and Liu, Yan-Rui and Zhu, Shi-Lin",
    title = "{An updated review of the new hadron states}",
    eprint = "2204.02649",
    archivePrefix = "arXiv",
    primaryClass = "hep-ph",
    doi = "10.1088/1361-6633/aca3b6",
    journal = "Rept. Prog. Phys.",
    volume = "86",
    number = "2",
    pages = "026201",
    year = "2023"
}

@article{Du:2022jjv,
    author = "Du, Meng-Lin and Albaladejo, Miguel and Guo, Feng-Kun and Nieves, Juan",
    title = "{Combined analysis of the $Z_c(3900)$ and the $Z_{cs}(3985)$ exotic states}",
    eprint = "2201.08253",
    archivePrefix = "arXiv",
    primaryClass = "hep-ph",
    doi = "10.1103/PhysRevD.105.074018",
    journal = "Phys. Rev. D",
    volume = "105",
    number = "7",
    pages = "074018",
    year = "2022"
}

@article{Yang:2020nrt,
    author = "Yang, Zhi and Cao, Xu and Guo, Feng-Kun and Nieves, Juan and Valderrama, Manuel Pavon",
    title = "{Strange molecular partners of the $Z_c(3900)$ and $Z_c(4020)$}",
    eprint = "2011.08725",
    archivePrefix = "arXiv",
    primaryClass = "hep-ph",
    doi = "10.1103/PhysRevD.103.074029",
    journal = "Phys. Rev. D",
    volume = "103",
    number = "7",
    pages = "074029",
    year = "2021"
}

@article{Ortega:2021enc,
    author = "Ortega, Pablo G. and Entem, David R. and Fernandez, F.",
    title = "{The strange partner of the $Z_c$ structures in a coupled-channels model}",
    eprint = "2103.07871",
    archivePrefix = "arXiv",
    primaryClass = "hep-ph",
    doi = "10.1016/j.physletb.2021.136382",
    journal = "Phys. Lett. B",
    volume = "818",
    pages = "136382",
    year = "2021"
}

@article{Meng:2021rdg,
    author = "Meng, Lu and Wang, Bo and Wang, Guang-Juan and Zhu, Shi-Lin",
    title = "{Implications of the $Z_{cs}(3985)$ and $Z_{cs}(4000)$ as two different states}",
    eprint = "2104.08469",
    archivePrefix = "arXiv",
    primaryClass = "hep-ph",
    doi = "10.1016/j.scib.2021.06.026",
    journal = "Sci. Bull.",
    volume = "66",
    pages = "2065--2071",
    year = "2021"
}

@article{Ikeno:2021mcb,
    author = "Ikeno, Natsumi and Molina, Raquel and Oset, Eulogio",
    title = "{$Z_{cs}$ states from the $D^*_s \bar{D}$ and $J/\psi K^*$ coupled channels: Signal in $B^+ \to J/\psi \tau K^*$ decay}",
    eprint = "2111.05024",
    archivePrefix = "arXiv",
    primaryClass = "hep-ph",
    doi = "10.1103/PhysRevD.105.014012",
    journal = "Phys. Rev. D",
    volume = "105",
    number = "1",
    pages = "014012",
    year = "2022",
    note = "[Erratum: Phys.Rev.D 106, 099905 (2022)]"
}

@article{Meng:2020ihj,
    author = "Meng, Lu and Wang, Bo and Zhu, Shi-Lin",
    title = "{$Z_{cs}(3985)^-$ as the $U$-spin partner of $Z_c(3900)^-$ and implication of other states in the $\text{SU(3)}_F$ symmetry and heavy quark symmetry}",
    eprint = "2011.08656",
    archivePrefix = "arXiv",
    primaryClass = "hep-ph",
    doi = "10.1103/PhysRevD.102.111502",
    journal = "Phys. Rev. D",
    volume = "102",
    number = "11",
    pages = "111502",
    year = "2020"
}

@misc{Du:2020vwb,
    author = "Du, Meng-Chuan and Wang, Qian and Zhao, Qiang",
    title = "{The nature of charged charmonium-like states $Z_c(3900)$ and its strange partner $Z_{cs}(3982)$}",
    eprint = "2011.09225",
    archivePrefix = "arXiv",
    primaryClass = "hep-ph",
    month = "11",
    year = "2020"
}

@article{Yan:2021tcp,
    author = "Yan, Mao-Jun and Peng, Fang-Zheng and S{\'a}nchez S{\'a}nchez, Mario and Pavon Valderrama, Manuel",
    title = "{Axial meson exchange and the $Z_c(3900)$ and $Z_{cs}(3985)$ resonances as heavy hadron molecules}",
    eprint = "2102.13058",
    archivePrefix = "arXiv",
    primaryClass = "hep-ph",
    doi = "10.1103/PhysRevD.104.114025",
    journal = "Phys. Rev. D",
    volume = "104",
    number = "11",
    pages = "114025",
    year = "2021"
}

@article{Wang:2020iqt,
    author = "Wang, Zhi-Gang",
    title = "{Analysis of $Z_{cs}(3985)$ as the axialvector tetraquark state}",
    eprint = "2011.10959",
    archivePrefix = "arXiv",
    primaryClass = "hep-ph",
    doi = "10.1088/1674-1137/abfa83",
    journal = "Chin. Phys. C",
    volume = "45",
    number = "7",
    pages = "073107",
    year = "2021"
}

@article{Wan:2020oxt,
    author = "Wan, Bing-Dong and Qiao, Cong-Feng",
    title = "{About the exotic structure of $Z_{cs}$}",
    eprint = "2011.08747",
    archivePrefix = "arXiv",
    primaryClass = "hep-ph",
    doi = "10.1016/j.nuclphysb.2021.115450",
    journal = "Nucl. Phys. B",
    volume = "968",
    pages = "115450",
    year = "2021"
}

@article{Wang:2020kej,
    author = "Wang, Jun-Zhang and Zhou, Qin-Song and Liu, Xiang and Matsuki, Takayuki",
    title = "{Toward charged $Z_{cs}(3985)$ structure under a reflection mechanism}",
    eprint = "2011.08628",
    archivePrefix = "arXiv",
    primaryClass = "hep-ph",
    doi = "10.1140/epjc/s10052-021-08877-4",
    journal = "Eur. Phys. J. C",
    volume = "81",
    number = "1",
    pages = "51",
    year = "2021"
}

@article{2103.05282,
    author = "Ge, Ying-Hui and Liu, Xiao-Hai and Ke, Hong-Wei",
    title = "{Threshold effects as the origin of $Z_{cs}(4000)$, $Z_{cs}(4220)$ and $X(4700)$ observed in $B^+\to J/\psi \phi K^+$}",
    eprint = "2103.05282",
    archivePrefix = "arXiv",
    primaryClass = "hep-ph",
    doi = "10.1140/epjc/s10052-021-09590-y",
    journal = "Eur. Phys. J. C",
    volume = "81",
    number = "9",
    pages = "854",
    year = "2021"
}

@article{Belle:2008qeq,
    author = "Mizuk, R. and others",
    collaboration = "Belle",
    title = "{Observation of two resonance-like structures in the $\pi^+ \chi_{c1}$ mass distribution in exclusive $\bar{B}^0 \to K^- \pi^+ \chi_{c1}$ decays}",
    eprint = "0806.4098",
    archivePrefix = "arXiv",
    primaryClass = "hep-ex",
    reportNumber = "BELLE-CONF-0848",
    doi = "10.1103/PhysRevD.78.072004",
    journal = "Phys. Rev. D",
    volume = "78",
    pages = "072004",
    year = "2008"
}

@article{Ding:2008gr,
    author = "Ding, Gui-Jun",
    title = "{Are $Y(4260)$ and $Z^+_2(4250)$ $D_1 D$ or $D_0 D^*$ hadronic molecules?}",
    eprint = "0809.4818",
    archivePrefix = "arXiv",
    primaryClass = "hep-ph",
    doi = "10.1103/PhysRevD.79.014001",
    journal = "Phys. Rev. D",
    volume = "79",
    pages = "014001",
    year = "2009"
}

@article{Lee:2008tz,
    author = "Lee, Su Houng and Morita, Kenji and Nielsen, Marina",
    title = "{Width of exotics from QCD sum rules: Tetraquarks or molecules?}",
    eprint = "0808.3168",
    archivePrefix = "arXiv",
    primaryClass = "hep-ph",
    doi = "10.1103/PhysRevD.78.076001",
    journal = "Phys. Rev. D",
    volume = "78",
    pages = "076001",
    year = "2008"
}

@article{Wang:2007ea,
    author = "Wang, X. L. and others",
    collaboration = "Belle",
    title = "{Observation of Two Resonant Structures in $e^+e^- \to \pi^+ \pi^- \psi(2S)$ via Initial State Radiation at Belle}",
    eprint = "0707.3699",
    archivePrefix = "arXiv",
    primaryClass = "hep-ex",
    reportNumber = "BELLE-2007-33, KEK-2007-27",
    doi = "10.1103/PhysRevLett.99.142002",
    journal = "Phys. Rev. Lett.",
    volume = "99",
    pages = "142002",
    year = "2007"
}

@article{Ding:2007rg,
    author = "Ding, Gui-Jun and Zhu, Jie-Jie and Yan, Mu-Lin",
    title = "{Canonical Charmonium Interpretation for $Y(4360)$ and $Y(4660)$}",
    eprint = "0708.3712",
    archivePrefix = "arXiv",
    primaryClass = "hep-ph",
    doi = "10.1103/PhysRevD.77.014033",
    journal = "Phys. Rev. D",
    volume = "77",
    pages = "014033",
    year = "2008"
}

@article{Li:2009zu,
    author = "Li, Bai-Qing and Chao, Kuang-Ta",
    title = "{Higher Charmonia and $X$,$Y$,$Z$ states with Screened Potential}",
    eprint = "0903.5506",
    archivePrefix = "arXiv",
    primaryClass = "hep-ph",
    doi = "10.1103/PhysRevD.79.094004",
    journal = "Phys. Rev. D",
    volume = "79",
    pages = "094004",
    year = "2009"
}

@article{Guo:2008zg,
    author = "Guo, Feng-Kun and Hanhart, Christoph and Meissner, Ulf-G.",
    title = "{Evidence that the $Y(4660)$ is a $f_0(980) \psi^{\prime}$ bound state}",
    eprint = "0803.1392",
    archivePrefix = "arXiv",
    primaryClass = "hep-ph",
    reportNumber = "FZJ-IKP-TH-2008-05, HISKP-TH-08-05",
    doi = "10.1016/j.physletb.2008.05.057",
    journal = "Phys. Lett. B",
    volume = "665",
    pages = "26--29",
    year = "2008"
}

@article{Guo:2009id,
    author = "Guo, Feng-Kun and Hanhart, Christoph and Meissner, Ulf-G.",
    title = "{Implications of heavy quark spin symmetry on heavy meson hadronic molecules}",
    eprint = "0904.3338",
    archivePrefix = "arXiv",
    primaryClass = "hep-ph",
    reportNumber = "FZJ-IKP-TH-2009-13, HISKP-TH-09-16",
    doi = "10.1103/PhysRevLett.102.242004",
    journal = "Phys. Rev. Lett.",
    volume = "102",
    pages = "242004",
    year = "2009"
}

@article{Albuquerque:2011ix,
    author = "Albuquerque, Raphael M. and Nielsen, Marina and Rodrigues da Silva, Romulo",
    title = "{Exotic $1^{--}$ States in QCD Sum Rules}",
    eprint = "1110.2113",
    archivePrefix = "arXiv",
    primaryClass = "hep-ph",
    doi = "10.1103/PhysRevD.84.116004",
    journal = "Phys. Rev. D",
    volume = "84",
    pages = "116004",
    year = "2011"
}

@article{Badalian:2008dv,
    author = "Badalian, A. M. and Bakker, B. L. G. and Danilkin, I. V.",
    title = "{The $S$ - $D$ mixing and di-electron widths of higher charmonium $1^{--}$ states}",
    eprint = "0805.2291",
    archivePrefix = "arXiv",
    primaryClass = "hep-ph",
    doi = "10.1134/S1063778809040085",
    journal = "Phys. Atom. Nucl.",
    volume = "72",
    pages = "638--646",
    year = "2009"
}

@article{Simonov:2008cr,
    author = "Simonov, Yu. A. and Veselov, A. I.",
    title = "{Bottomonium $\Upsilon(5S)$ decays into $BB$ and $BB\pi$}",
    eprint = "0805.4518",
    archivePrefix = "arXiv",
    primaryClass = "hep-ph",
    doi = "10.1134/S002136400813002X",
    journal = "JETP Lett.",
    volume = "88",
    pages = "5--7",
    year = "2008"
}

@article{Dai:2017fwx,
    author = "Dai, Ling-Yun and Haidenbauer, Johann and Mei{\ss}ner, Ulf G.",
    title = "{Re-examining the $X(4630)$ resonance in the reaction $e^+e^-\to \Lambda^+_c\bar\Lambda^-_c$}",
    eprint = "1710.03142",
    archivePrefix = "arXiv",
    primaryClass = "hep-ph",
    doi = "10.1103/PhysRevD.96.116001",
    journal = "Phys. Rev. D",
    volume = "96",
    number = "11",
    pages = "116001",
    year = "2017"
}

@article{Ahmad:2025mue,
    author = "Ahmad, Zaki and Asghar, Ishrat and Akram, Faisal and Masud, Bilal",
    title = "{Strong decays of charmonia}",
    eprint = "2504.07605",
    archivePrefix = "arXiv",
    primaryClass = "hep-ph",
    doi = "10.1103/PhysRevD.111.034007",
    journal = "Phys. Rev. D",
    volume = "111",
    number = "3",
    pages = "034007",
    year = "2025"
}

@article{Maiani:2014aja,
    author = "Maiani, L. and Piccinini, F. and Polosa, A. D. and Riquer, V.",
    title = "{The $Z(4430)$ and a New Paradigm for Spin Interactions in Tetraquarks}",
    eprint = "1405.1551",
    archivePrefix = "arXiv",
    primaryClass = "hep-ph",
    doi = "10.1103/PhysRevD.89.114010",
    journal = "Phys. Rev. D",
    volume = "89",
    pages = "114010",
    year = "2014"
}

@article{Ebert:2008kb,
    author = "Ebert, D. and Faustov, R. N. and Galkin, V. O.",
    title = "{Excited heavy tetraquarks with hidden charm}",
    eprint = "0808.3912",
    archivePrefix = "arXiv",
    primaryClass = "hep-ph",
    doi = "10.1140/epjc/s10052-008-0754-8",
    journal = "Eur. Phys. J. C",
    volume = "58",
    pages = "399--405",
    year = "2008"
}

@article{Chen:2010ze,
    author = "Chen, Wei and Zhu, Shi-Lin",
    title = "{The Vector and Axial-Vector Charmonium-like States}",
    eprint = "1010.3397",
    archivePrefix = "arXiv",
    primaryClass = "hep-ph",
    doi = "10.1103/PhysRevD.83.034010",
    journal = "Phys. Rev. D",
    volume = "83",
    pages = "034010",
    year = "2011"
}

@article{Casalbuoni:1996pg,
    author = "Casalbuoni, R. and Deandrea, A. and Di Bartolomeo, N. and Gatto, Raoul and Feruglio, F. and Nardulli, G.",
    title = "{Phenomenology of heavy meson chiral Lagrangians}",
    eprint = "hep-ph/9605342",
    archivePrefix = "arXiv",
    reportNumber = "UGVA-DPT-1996-05-928, BARI-TH-96-237",
    doi = "10.1016/S0370-1573(96)00027-0",
    journal = "Phys. Rept.",
    volume = "281",
    pages = "145--238",
    year = "1997"
}

@article{Colangelo:2012xi,
    author = "Colangelo, P. and De Fazio, F. and Giannuzzi, F. and Nicotri, S.",
    title = "{New meson spectroscopy with open charm and beauty}",
    eprint = "1207.6940",
    archivePrefix = "arXiv",
    primaryClass = "hep-ph",
    reportNumber = "BARI-TH-651-12",
    doi = "10.1103/PhysRevD.86.054024",
    journal = "Phys. Rev. D",
    volume = "86",
    pages = "054024",
    year = "2012"
}

@article{Weinberg:1962hj,
    author = "Weinberg, Steven",
    title = "{Elementary particle theory of composite particles}",
    doi = "10.1103/PhysRev.130.776",
    journal = "Phys. Rev.",
    volume = "130",
    pages = "776--783",
    year = "1963"
}

@article{Meissner:2015mza,
    author = "Mei{\ss}ner, Ulf-G. and Oller, Jos{\'e} A.",
    title = "{Testing the $\chi_{c1}\, p$ composite nature of the $P_c(4450)$}",
    eprint = "1507.07478",
    archivePrefix = "arXiv",
    primaryClass = "hep-ph",
    doi = "10.1016/j.physletb.2015.10.015",
    journal = "Phys. Lett. B",
    volume = "751",
    pages = "59--62",
    year = "2015"
}

@article{Guo:2019kdc,
    author = "Guo, Zhi-Hui and Oller, J. A.",
    title = "{Anatomy of the newly observed hidden-charm pentaquark states: $P_c(4312)$, $P_c(4440)$ and $P_c(4457)$}",
    eprint = "1904.00851",
    archivePrefix = "arXiv",
    primaryClass = "hep-ph",
    doi = "10.1016/j.physletb.2019.04.053",
    journal = "Phys. Lett. B",
    volume = "793",
    pages = "144--149",
    year = "2019"
}

@article{Oller:2000fj,
    author = "Oller, J. A. and Meissner, Ulf G.",
    title = "{Chiral dynamics in the presence of bound states: Kaon nucleon interactions revisited}",
    eprint = "hep-ph/0011146",
    archivePrefix = "arXiv",
    reportNumber = "FZJ-IKP-TH-2000-26",
    doi = "10.1016/S0370-2693(01)00078-8",
    journal = "Phys. Lett. B",
    volume = "500",
    pages = "263--272",
    year = "2001"
}

@article{Oller:2006jw,
    author = "Oller, Jose A.",
    title = "{On the strangeness -1 $S$-wave meson-baryon scattering}",
    eprint = "hep-ph/0603134",
    archivePrefix = "arXiv",
    doi = "10.1140/epja/i2006-10011-3",
    journal = "Eur. Phys. J. A",
    volume = "28",
    pages = "63--82",
    year = "2006"
}

@article{Guo:2020vmu,
    author = "Guo, Zhi-Hui and Oller, J. A.",
    title = "{Unified description of the hidden-charm tetraquark states $Z_{cs}(3985), Z_c(3900)$, and $X(4020)$}",
    eprint = "2012.11904",
    archivePrefix = "arXiv",
    primaryClass = "hep-ph",
    doi = "10.1103/PhysRevD.103.054021",
    journal = "Phys. Rev. D",
    volume = "103",
    number = "5",
    pages = "054021",
    year = "2021"
}

@article{Kang:2016ezb,
    author = "Kang, Xian-Wei and Guo, Zhi-Hui and Oller, J. A.",
    title = "{General considerations on the nature of $Z_b(10610)$ and $Z_b(10650)$ from their pole positions}",
    eprint = "1603.05546",
    archivePrefix = "arXiv",
    primaryClass = "hep-ph",
    doi = "10.1103/PhysRevD.94.014012",
    journal = "Phys. Rev. D",
    volume = "94",
    number = "1",
    pages = "014012",
    year = "2016"
}

@article{Zou:2002ar,
    author = "Zou, B. S. and Bugg, D. V.",
    title = "{Covariant tensor formalism for partial wave analyses of psi decay to mesons}",
    eprint = "hep-ph/0211457",
    archivePrefix = "arXiv",
    doi = "10.1140/epja/i2002-10135-4",
    journal = "Eur. Phys. J. A",
    volume = "16",
    pages = "537--547",
    year = "2003"
}

@article{Dulat:2005in,
    author = "Dulat, Sayipjamal and Zou, Bing-Song",
    title = "{Covariant tensor formalism for partial wave analyses of $\psi$ decays into $\gamma B\bar B$, $\gamma\gamma V$ and $\psi(2S)\to\gamma\chi_{c0,1,2}$ with $\chi_{c0,1,2}\to K\bar K \pi^+\pi^- $ and $2\pi^+2\pi^-$}",
    eprint = "hep-ph/0508087",
    archivePrefix = "arXiv",
    doi = "10.1140/epja/i2005-10140-1",
    journal = "Eur. Phys. J. A",
    volume = "26",
    pages = "125--134",
    year = "2005",
    note = "[Erratum: Eur.Phys.J.A 56, 275 (2020)]"
}

@misc{Huang:2026egv,
    author = "Huang, Hong and Wang, Yi-Ning and Yu, Jiang-Hao",
    title = "{Covariant canonical-spinor amplitudes for partial wave analysis}",
    eprint = "2603.04487",
    archivePrefix = "arXiv",
    primaryClass = "hep-ph",
    month = "3",
    year = "2026"
}

@article{Liu:2024tgq,
    author = "Liu, Cheng-Xi and Man, Zi-Long and Gao, Tian-Le and Liu, Xiang",
    title = "{Prospects for observing the missing $2D$ and $1F$ charmonium states around 4~GeV}",
    eprint = "2411.15689",
    archivePrefix = "arXiv",
    primaryClass = "hep-ph",
    doi = "10.1103/lgfd-3pqc",
    journal = "Phys. Rev. D",
    volume = "113",
    number = "7",
    pages = "074009",
    year = "2026"
}

@article{Gortchakov:1995im,
    author = "Gortchakov, O. and Locher, M. P. and Markushin, V. E. and von Rotz, S.",
    title = "{Two meson doorway calculation for $\bar{p} p\to \phi \pi$ including off-shell effects and the OZI rule}",
    reportNumber = "PSI-PR-95-15, ZU-TH-17-95",
    doi = "10.1007/BF01285155",
    journal = "Z. Phys. A",
    volume = "353",
    pages = "447--453",
    year = "1996"
}

@article{ParticleDataGroup:2024cfk,
    author = "Navas, S. and others",
    collaboration = "Particle Data Group",
    title = "{Review of particle physics}",
    doi = "10.1103/PhysRevD.110.030001",
    journal = "Phys. Rev. D",
    volume = "110",
    number = "3",
    pages = "030001",
    year = "2024"
}

@article{Wang:2019mhs,
    author = "Wang, Jun-Zhang and Chen, Dian-Yong and Liu, Xiang and Matsuki, Takayuki",
    title = "{Constructing $J/\psi$ family with updated data of charmoniumlike $Y$ states}",
    eprint = "1903.07115",
    archivePrefix = "arXiv",
    primaryClass = "hep-ph",
    doi = "10.1103/PhysRevD.99.114003",
    journal = "Phys. Rev. D",
    volume = "99",
    number = "11",
    pages = "114003",
    year = "2019"
}

@article{Cheng:2004ru,
    author = "Cheng, Hai-Yang and Chua, Chun-Khiang and Soni, Amarjit",
    title = "{Final state interactions in hadronic $B$ decays}",
    eprint = "hep-ph/0409317",
    archivePrefix = "arXiv",
    reportNumber = "BNL-HET-04-17",
    doi = "10.1103/PhysRevD.71.014030",
    journal = "Phys. Rev. D",
    volume = "71",
    pages = "014030",
    year = "2005"
}

@article{Wu:2019vbk,
    author = "Wu, Qi and Chen, Dian-Yong and Fan, Xue-Jia and Li, Gang",
    title = "{Production of $Z_c(3900)$ and $Z_c(4020)$ in $B_c$ decay}",
    eprint = "1902.05737",
    archivePrefix = "arXiv",
    primaryClass = "hep-ph",
    doi = "10.1140/epjc/s10052-019-6784-6",
    journal = "Eur. Phys. J. C",
    volume = "79",
    number = "3",
    pages = "265",
    year = "2019"
}

@article{Guo:2010ak,
    author = "Guo, Feng-Kun and Hanhart, Christoph and Li, Gang and Meissner, Ulf-G. and Zhao, Qiang",
    title = "{Effect of charmed meson loops on charmonium transitions}",
    eprint = "1008.3632",
    archivePrefix = "arXiv",
    primaryClass = "hep-ph",
    reportNumber = "FZJ-IKP-TH-2010-08, HISKP-TH-10-09",
    doi = "10.1103/PhysRevD.83.034013",
    journal = "Phys. Rev. D",
    volume = "83",
    pages = "034013",
    year = "2011"
}

@article{Liu:2013vfa,
    author = "Liu, Xiao-Hai and Li, Gang",
    title = "{Exploring the threshold behavior and implications on the nature of $Y(4260)$ and $Z_c(3900)$}",
    eprint = "1306.1384",
    archivePrefix = "arXiv",
    primaryClass = "hep-ph",
    doi = "10.1103/PhysRevD.88.014013",
    journal = "Phys. Rev. D",
    volume = "88",
    pages = "014013",
    year = "2013"
}

@article{Wang:2025dur,
    author = "Wang, Xiongfei and Liu, Xiang and Gao, Yuanning",
    title = "{Colloquium: Hadron production in open-charm meson pairs at $e^+e^-$ colliders}",
    eprint = "2502.15117",
    archivePrefix = "arXiv",
    primaryClass = "hep-ex",
    doi = "10.1103/2mrp-chly",
    journal = "Rev. Mod. Phys.",
    volume = "98",
    number = "2",
    pages = "021001",
    year = "2026"
}
\end{document}